\documentclass[
  aps,
  prb,
  twocolumn,
  superscriptaddress,
  longbibliography,
  amsmath,
  amssymb,
  nofootinbib
]{revtex4-2}

\usepackage{amsfonts}
\usepackage{bm}
\usepackage{mathtools}
\usepackage{physics}
\usepackage{microtype}
\usepackage{booktabs}
\usepackage{hyperref}

\hypersetup{
  colorlinks=true,
  linkcolor=blue,
  citecolor=blue,
  urlcolor=cyan,
  pdftitle={Long-range Nonlinear Sigma Model for a Singular Quantum Kicked Rotor},
  pdfauthor={Weitao Chen and Yunxiang Liao}
}

\allowdisplaybreaks[2]
\newcommand{\iu}{\mathrm{i}}
\newcommand{\e}{\mathrm{e}}
\newcommand{\str}{\operatorname{str}}
\newcommand{\Str}{\operatorname{Str}}
\newcommand{\sdet}{\operatorname{sdet}}
\newcommand{\Sdet}{\operatorname{Sdet}}
\newcommand{\cZ}{\mathcal Z}

\newcommand{\cS}{\mathcal S}

\newcommand{\AR}{\mathrm{AR}}
\newcommand{\BF}{\mathrm{BF}}
\newcommand{\QD}{\mathrm{QD}}
\newcommand{\hbareff}{\hbar}
\newcommand{\av}[1]{\left\langle #1\right\rangle}
\begin{document}

\title{Long-range Nonlinear Sigma Model for a Singular Quantum Kicked Rotor}

\author{Weitao Chen}
\email{weitao.chen.1@warwick.ac.uk}
\affiliation{Department of Physics, University of Warwick, Coventry, CV4 7AL, United Kingdom}
\author{Yunxiang Liao}
\email{yunlia@kth.se}
\affiliation{Department of Physics, KTH Royal Institute of Technology, SE-106 91 Stockholm, Sweden}
\begin{abstract}
Singular kicked rotors have long been compared with power-law random banded matrices (PRBM) because their momentum-space Floquet matrix elements decay algebraically. However, it has remained unclear whether the deterministic correlations of the rotor become irrelevant at long distances and, consequently, under what conditions the two systems share the same infrared theory. To address this question, we derive a nonlocal supersymmetric nonlinear sigma model directly from a quantum kicked rotor with a power-law or logarithmic singularity. By carrying out the renormalization-group analysis up to two-loop order, we show that, after matching the symmetry class and coupling convention, the rotor reproduces the long-range Anderson transition of the corresponding PRBM, including its localized, critical, and extended infrared regimes.
\end{abstract}
\

\maketitle

\section{Introduction}
\label{sec:introduction}

The quantum kicked rotor is a paradigmatic model of quantum chaos.  Its
classical dynamics can exhibit unbounded diffusive growth in momentum,
whereas quantum interference suppresses this diffusion and produces
dynamical localization.  The classical model originates from the standard
map and the resonance-overlap picture introduced by Chirikov
\cite{Chirikov1979}, while its quantum version became a simple but powerful
setting for studying spectral statistics, quantum--classical
correspondence, and dynamical localization
\cite{Izrailev1990,BenentiCasatiGongZou2026}.  The resulting localization
is the momentum-space analogue of Anderson localization in a
one-dimensional disordered lattice
\cite{Anderson1958,AbrahamsBook2010,EversMirlin2008}.  One important reason
for the broad use of the kicked rotor is that its periodically driven form
is considerably easier to implement experimentally than a microscopic
disordered lattice with the same effective physics
\cite{MooreEtAl1995}.

On the theoretical side, the connection between the kicked rotor and the
Anderson problem has been established at several complementary levels.  The
Floquet eigenvalue equation can be mapped onto a one-dimensional
tight-binding problem in momentum space with pseudorandom onsite phases
\cite{FishmanGrempelPrange1982,GrempelPrangeFishman1984}.  Diagrammatic
methods identify the interference corrections responsible for dynamical
localization~\cite{Altland1993}, while the supersymmetric formulation yields
the same long-wavelength nonlinear sigma model as the corresponding
quasi-one-dimensional Anderson system
\cite{AltlandZirnbauerQKR1996}.  The latter result is especially important:
it demonstrates that localization phenomena normally associated with
quenched disorder can be generated dynamically by a single deterministic
Floquet operator.  The incommensurate free-rotation phases play the role of
pseudorandom onsite phases, while no external random potential is added \cite{Birkhoff1931}.

Along this idea, several variants of the kicked rotor have been introduced
to reproduce Anderson transitions that are absent in the conventional
one-dimensional short-range problem and therefore require additional effective
dimensionality.  One possibility is
to add physical or synthetic dimensions.  The original quasiperiodically
driven construction showed that a one-dimensional rotor with additional
incommensurate frequencies can reproduce a higher-dimensional Anderson
problem~\cite{CasatiGuarneriShepelyansky1989}.  A genuinely
three-dimensional kicked rotor was also shown to possess critical spectral and dynamical behavior~\cite{WangGarciaGarcia2009}.
The quasiperiodic atomic kicked rotor subsequently enabled the experimental
observation of the three-dimensional Anderson transition and the finite-time scaling determination of its critical
exponent~\cite{ChabeEtAl2008,LemarieGremaudDelande2009,LemarieEtAl2010PRL}.  Its microscopic
field theory further clarified how the synthetic dimensions and arithmetic
properties of the effective Planck constant determine the universality
class~\cite{TianAltlandGarst2011,TianAltland2012}.  Kicked dynamics on
small-world graphs extends this strategy to the critical dynamics of the
effectively infinite-dimensional Anderson transition
\cite{ChenEtAlSmallWorld2026}.

A distinct route to an Anderson transition in one dimension is to
introduce long-range transitions through a singular kicking potential.  A
nonanalytic kick changes the ordinary kicked-rotor problem qualitatively:
it generates algebraically decaying momentum-space Floquet matrix elements
and anomalous classical transport \cite{wang2022statistical}.  For a logarithmic singularity,
numerical work found multifractal Floquet eigenstates and spectral
statistics intermediate between the Wigner--Dyson and Poisson limits
\cite{GarciaGarciaWang2005}.  More generally, singular long-range hopping
produces anomalous wave-packet dynamics whose time and finite-size scaling
are closely connected to multifractality
\cite{ChenLemarieGong2023}.  These results naturally suggest a connection
with power-law random banded matrices (PRBM), which provide a
one-dimensional model of the long-range Anderson transition.

In the conventional PRBM ensemble, the variance of the matrix elements
decays algebraically away from the diagonal. If the standard deviation of the
hopping amplitude behaves as
\(
  [\operatorname{Var}(H_{nm})]^{1/2}\sim |n-m|^{-\alpha},
\)
the standard ensemble is extended for \(\alpha<1\), localized for
\(\alpha>1\), and critical at \(\alpha=1\)
\cite{MirlinEtAl1996,fan2026localization}.  At the critical exponent, the
eigenstates are multifractal and the level statistics interpolate between
the random-matrix and Poisson limits
\cite{MirlinEvers2000,EversMirlin2008}.  The corresponding nonlocal
nonlinear sigma model gives controlled predictions for the
renormalization-group flow~\cite{MirlinEtAl1996}, participation moments and
eigenfunction correlations
\cite{FyodorovMirlin1995,MirlinEvers2000}, and spectral statistics
\cite{MirlinReview2000}.  PRBM therefore provides not only a qualitative
analogy but also a detailed field-theory benchmark for the singular kicked
rotor.

The PRBM idea has also been extended to other long-range random-matrix
ensembles.  In particular, modified variance profiles can produce
logarithmic rather than ordinary power-law multifractality
\cite{ChenEtAl2024PRR}.  Such ensembles can reproduce the strong finite-size
effects and critical localization associated with Anderson transitions in
effectively infinite dimension~\cite{ChenEtAl2024PRB}. These examples illustrate that long-range random-matrix behavior is sensitive
to the detailed envelope and statistical structure of the matrix elements:
modifying the variance profile changes the long-distance hopping kernel and
can produce qualitatively different critical scaling.  Consequently, the
presence of algebraically decaying matrix elements alone is not sufficient to
establish a universality class; both their envelope and their statistical
correlations must be controlled.

This point is particularly important for the kicked rotor.  A conventional
PRBM contains quenched random diagonal and off-diagonal matrix elements,
whereas the rotor is represented by a single deterministic unitary matrix.
The incommensurate kinetic phases can be understood as dynamically
generated pseudorandom diagonal phases, but the off-diagonal matrix elements
are Fourier coefficients of the same kicking function and are therefore
strongly correlated and nonrandom.  Such correlations can qualitatively
change a long-range hopping problem.  In particular, correlation-induced
localization may occur even in a parameter regime where the corresponding
uncorrelated PRBM would be extended
\cite{NosovKhaymovichKravtsov2019}.  Therefore, the similarity of the
power-law decay alone does not prove that the singular rotor and PRBM share
the same eigenfunction statistics or the same infrared fixed point.

The central question is whether these deterministic Floquet correlations
become irrelevant after coarse graining, leaving the same nonlocal sigma
model as PRBM, or whether they generate additional long-distance
structures.  To answer this question, we derive a supersymmetric
nonlinear sigma model microscopically from the singular quantum kicked
rotor. We find that, after matching the symmetry class, coupling convention, source
normalization, and asymptotic scaling window, the localized and critical
regimes reproduce the corresponding class-A PRBM theory.  The localized
phase develops the same localization-volume description, while the
logarithmic rotor realizes the marginal critical sigma model with
multifractal eigenstates and critical spectral statistics.  The extended
phase shares the same class-A metallic infrared fixed point, but its
approach to this limit is more subtle.  The divergence of the singular
potential, the complete deterministic relaxation kernel, and the
competition between endpoint and stationary-phase asymptotics introduce
additional finite-scale structures.  The derivation therefore determines
both when PRBM universality applies and which dynamical corrections remain
specific to the kicked rotor.

The remainder of the paper is organized as follows.
Section~\ref{sec:model} defines the class-A singular rotor, its smooth inversion-breaking harmonic, and its regularization.  Section~\ref{sec:microscopic} constructs a common
supersymmetric generating functional for the inverse participation ratios
and the two-level quasienergy correlator using the color--flavor
transformation.  Section~\ref{sec:fourier} derives the nonlocal kernel from
the one-kick probability and determines the relevant asymptotic and
finite-size windows.  Section~\ref{sec:predictions} performs the
background-field expansion and the phase-dependent renormalization-group
analysis and evaluates the IPR and spectral observables.
Section~\ref{sec:universality} discusses the universality and limitations of the proposed sigma model and
the RG analysis.
Finally, Sec.~\ref{sec:conclusions} summarizes the main results.

\section{Singular Kicked Rotor Model}
\label{sec:model}

This section introduces the deterministic Floquet operator and the scales that
enter the field theory.  The kicked-rotor representation and its relation to
momentum-space localization follow the standard construction of
Refs.~\cite{FishmanGrempelPrange1982,GrempelPrangeFishman1984,
Altland1993,AltlandZirnbauerQKR1996}.  We work in class A from the beginning,
using the Wigner--Dyson and nonlinear-sigma-model classifications~\cite{Dyson1962,AltlandZirnbauer1997,EversMirlin2008}, so no
orthogonal-class
zero-mode factors or Cooperon contractions appear later.
Consider
\begin{equation}
  \widehat{\mathcal H}(t)
  =
  \frac{\widehat p^{\,2}}{2}
  +
  K V_{\beta,\eta}(\widehat x)
  \sum_{m\in\mathbb Z}\delta(t-m),
  \qquad
  [\widehat x,\widehat p]=\iu\hbareff.
  \label{eq:H}
\end{equation}
The rotor lives on \(x\in[-\pi,\pi)\).  We introduce a generic dimensionless
flux \(\chi\) through
\begin{equation}
  \widehat p=\hbareff(\widehat n+\chi),
  \qquad
  \chi\notin\left\{0,\frac12\right\}
  \quad (\mathrm{mod}\ 1),
  \label{eq:flux}
\end{equation}
which breaks the ordinary spinless time-reversal symmetry that reverses
momentum.  The Hilbert-space coordinate \(n\in\mathbb Z\) will become the
spatial coordinate of the one-dimensional sigma model.  We introduce
\begin{equation}
  \kappa\equiv\frac{K}{\hbareff}
  \label{eq:kappa}
\end{equation}
because the kick operator depends on \(K\) and \(\hbareff\) through this
ratio.

To periodize the singularity without creating a cusp at the boundary, set
\begin{equation}
  s(x)=2\left|\sin\frac{x}{2}\right|,
  \label{eq:sx}
\end{equation}
and define the class-A kicking potential directly as
\begin{equation}
  V_{\beta,\eta}(x)
  =
  \begin{cases}
    [s(x)^\beta-1]/\beta+\eta\sin x,
      & \beta\neq0,\\[1mm]
    \ln s(x)+\eta\sin x,
      & \beta=0,
  \end{cases}
  \qquad \eta\neq0.
  \label{eq:V}
\end{equation}
Since \(s(x)=|x|+O(|x|^3)\) and \(\sin x=x+O(x^3)\), the odd harmonic is
smooth at the singular point and does not change its leading nonanalyticity.
It does, however, remove the residual inversion-assisted antiunitary symmetry
of the even singular kick.  More explicitly, a generic
\(V_{\beta,\eta}(-x)\neq V_{\beta,\eta}(x)\) makes the kick matrix
nonsymmetric in momentum space, while Eq.~\eqref{eq:flux} independently
breaks the ordinary momentum-reversing time reversal.  For generic
\(\chi\) and \(\eta\) no antiunitary symmetry remains, so the Floquet problem
is in class A.  We regard \(\eta\) as a fixed model parameter and suppress it
in the notation for the Floquet operator below.  The singular part of Eq.~\eqref{eq:V} is the periodic version
of the nonanalytic kicks considered in
Ref.~\cite{GarciaGarciaWang2005, ChenLemarieGong2023}.

A rounded core of angular size \(\delta_x\) is obtained by replacing \(s(x)\)
in Eq.~\eqref{eq:V} with
\begin{equation}
  s_\delta(x)
  =
  \left[
    4\sin^2\frac{x}{2}+\delta_x^2
  \right]^{1/2}.
  \label{eq:sdelta}
\end{equation}
The smooth term \(\eta\sin x\) needs no regularization.  For fixed
\(\delta_x>0\), algebraic Fourier tails terminate at
\begin{equation}
  r_\delta\sim\delta_x^{-1}.
  \label{eq:rdelta}
\end{equation}
A finite system of momentum length \(N\) probes the fractional infrared
regime only if \(N\delta_x\ll1\).

Using the state immediately after each kick, the Floquet operator is
\begin{equation}
  U_\chi
  =
  \exp\left[
    -\frac{\iu\hbareff}{2}
    (\widehat n+\chi)^2
  \right]
  \exp[-\iu\kappa V_{\beta,\eta}(\widehat x)].
  \label{eq:U}
\end{equation}
Its momentum-space matrix elements are
\begin{align}
  (U_\chi)_{nm}
  &=
  \e^{-\iu\hbareff(n+\chi)^2/2}
  u_{n-m},
  \label{eq:Unm}\\
  u_r
  &=
  \int_{-\pi}^{\pi}
  \frac{\dd x}{2\pi}\,
  \e^{-\iu\kappa V_{\beta,\eta}(x)}
  \e^{\iu r x}.
  \label{eq:ur}
\end{align}
The flux changes the kinetic phases but not \(|(U_\chi)_{nm}|^2\).
We also require \(\hbareff/(4\pi)\) to be irrational, or a rational
approximant whose resonance denominator exceeds every retained length.
These assumptions define a class-A, off-resonant deterministic rotor; the
importance of the arithmetic resonance denominator and symmetry sector in
kicked-rotor field theory is discussed in
Refs.~\cite{AltlandZirnbauerQKR1996,TianAltland2010,AltlandEtAl2015}.  We next
construct the source functional for eigenfunction and spectral
observables.
\section{generating functional for the action, IPR, and two-level correlator}
\label{sec:microscopic}

The supersymmetric generating-function strategy follows the
Wegner--Efetov formulation of localization~\cite{Wegner1979,Efetov1983,
EfetovBook}.  For unitary maps, the exact color--flavor transformation and
its application to circular ensembles and quantum-chaotic maps were
developed in Refs.~\cite{ZirnbauerCircular1996,ZirnbauerCFT1998,
ZirnbauerPair1999,ZirnbauerCFT2021}; the kicked-rotor implementation is
reviewed in Ref.~\cite{AltlandEtAl2015}.  We restate the construction here
because both source sectors must be normalized in the same convention. Global source
parameters generate the two-level quasienergy correlator, while local source
derivatives generate wave-function moments.
\subsection{The wave-function moments}
\label{subsec:ipr_definition}

Let
\begin{equation}
  U_\chi\ket{\mu}
  =
  \e^{-\iu\varphi_\mu}\ket{\mu},
  \qquad
  \psi_\mu(n)=\braket{n}{\mu}.
  \label{eq:eigenphase}
\end{equation}
The IPR convention used below is
\begin{align}
  I_p^{(\mu)}
  &\equiv
  \sum_{n=1}^{N}|\psi_\mu(n)|^{2p},
  \label{eq:Ip_state}\\
  \av{I_p}
  &\equiv
  \frac1N
  \sum_{\mu=1}^{N}I_p^{(\mu)}.
  \label{eq:Ip_average}
\end{align}
The brackets in Eq.~\eqref{eq:Ip_average} denote a spectral average over all
Floquet eigenstates, not a disorder average.  Local moments and IPR
operators in nonlinear sigma models were developed in
Refs.~\cite{HofWegner1986,Wegner1987I,Wegner1987II,FyodorovMirlin1995,
MirlinReview2000}.  The ordinary inverse
participation ratio is \(\av{I_2}\).  We also define the local moment
\begin{equation}
  M_p(n)
  \equiv
  \frac1N
  \sum_{\mu=1}^{N}
  |\psi_\mu(n)|^{2p},
  \qquad
  \av{I_p}=\sum_n M_p(n).
  \label{eq:Mp}
\end{equation}
This definition is used in the microscopic source formula, in the RG
calculation, and in all final predictions.

\subsection{Floquet resolvents and the equal-state pole}
\label{subsec:resolvents}

For \(a=\e^{-\gamma}\), with \(\gamma\to0^+\) an infinitesimal positive
regulator, define
\begin{align}
  G^+(a,\phi)
  &=
  [1-a\e^{\iu\phi}U_\chi]^{-1},
  \label{eq:Gplus}\\
  G^-(a,\phi)
  &=
  [1-a\e^{-\iu\phi}U_\chi^\dagger]^{-1}.
  \label{eq:Gminus}
\end{align}
We write
\begin{equation}
  \epsilon\equiv1-a^2=2\gamma+O(\gamma^2)
  \label{eq:epsilon}
\end{equation}
for the retarded--advanced regulator.  The diagonal spectral representations,
with \(\psi_\mu(n)=\braket{n}{\mu}\) defined in
Eq.~\eqref{eq:eigenphase}, are
\begin{align}
  G^+_{nn}(a,\phi)
  &=
  \sum_\mu
  \frac{|\psi_\mu(n)|^2}
  {1-a\e^{\iu(\phi-\varphi_\mu)}},
  \label{eq:Gplus_spectral}\\
  G^-_{nn}(a,\phi)
  &=
  \sum_\mu
  \frac{|\psi_\mu(n)|^2}
  {1-a\e^{-\iu(\phi-\varphi_\mu)}}.
  \label{eq:Gminus_spectral}
\end{align}
To evaluate their product, introduce
\begin{align}
  \mathcal I_{\mu_1\ldots\mu_p}(a)
  &\equiv
  \int_0^{2\pi}\frac{\dd\phi}{2\pi}
  \prod_{j=1}^{p-1}
  \frac{1}{1-a\e^{\iu(\phi-\varphi_{\mu_j})}}
  \nonumber\\
  &\qquad\times
  \frac{1}{1-a\e^{-\iu(\phi-\varphi_{\mu_p})}}.
  \label{eq:I_multiplet_definition}
\end{align}
For \(0<a<1\), every denominator is absolutely convergent as a geometric
series.  The retarded factors and the single advanced factor are
\begin{align}
  \frac{1}{1-a\e^{\iu(\phi-\varphi_{\mu_j})}}
  &=
  \sum_{k_j=0}^{\infty}
  a^{k_j}\e^{\iu k_j\phi}
  \e^{-\iu k_j\varphi_{\mu_j}},
  \label{eq:retarded_geometric_series}\\
  \frac{1}{1-a\e^{-\iu(\phi-\varphi_{\mu_p})}}
  &=
  \sum_{\ell=0}^{\infty}
  a^{\ell}\e^{-\iu\ell\phi}
  \e^{\iu\ell\varphi_{\mu_p}}.
  \label{eq:advanced_geometric_series}
\end{align}
Substitution into Eq.~\eqref{eq:I_multiplet_definition} gives
\begin{align}
  \mathcal I_{\mu_1\ldots\mu_p}(a)
  &=
  \sum_{k_1,\ldots,k_{p-1}\geq0}
  \sum_{\ell\geq0}
  a^{\sum_jk_j+\ell}
  \e^{-\iu\sum_jk_j\varphi_{\mu_j}}
  \e^{\iu\ell\varphi_{\mu_p}}
  \nonumber\\
  &\quad\times
  \int_0^{2\pi}\frac{\dd\phi}{2\pi}
  \e^{\iu(\sum_jk_j-\ell)\phi}.
  \label{eq:I_before_phi_integral}
\end{align}
The phase integral is the Fourier orthogonality relation
\begin{equation}
  \int_0^{2\pi}\frac{\dd\phi}{2\pi}
  \e^{\iu M\phi}
  =\delta_{M,0},
  \qquad M\in\mathbb Z.
  \label{eq:phi_orthogonality}
\end{equation}
It therefore imposes the integer constraint
\begin{equation}
  \ell=k_1+\cdots+k_{p-1}.
  \label{eq:advanced_index_constraint}
\end{equation}
Eliminating \(\ell\) and rearranging the phase factors gives
\begin{align}
  \mathcal I_{\mu_1\ldots\mu_p}(a)
  &=
  \sum_{k_1,\ldots,k_{p-1}\geq0}
  \prod_{j=1}^{p-1}
  \left[
    a^2\e^{-\iu(
      \varphi_{\mu_j}-\varphi_{\mu_p}
    )}
  \right]^{k_j}
  \nonumber\\
  &=
  \prod_{j=1}^{p-1}
  \sum_{k_j=0}^{\infty}
  \left[
    a^2\e^{-\iu(
      \varphi_{\mu_j}-\varphi_{\mu_p}
    )}
  \right]^{k_j}
  \nonumber\\
  &=
  \prod_{j=1}^{p-1}
  \frac{1}{
    1-a^2\e^{-\iu(
      \varphi_{\mu_j}-\varphi_{\mu_p}
    )}
  }.
  \label{eq:I_multiplet_evaluated}
\end{align}
Thus the \(\phi\) integration is not an approximation: it exactly equates
the advanced winding number to the sum of all retarded winding numbers.
Consequently,
\begin{align}
  &\int_0^{2\pi}\frac{\dd\phi}{2\pi}
  [G^+_{nn}(a,\phi)]^{p-1}
  G^-_{nn}(a,\phi)
  \nonumber\\
  &\quad=
  \sum_{\mu_1,\ldots,\mu_p}
  \left[
    \prod_{j=1}^{p}
    |\psi_{\mu_j}(n)|^2
  \right]
  \mathcal I_{\mu_1\ldots\mu_p}(a).
  \label{eq:resolvent_product_spectral}
\end{align}
For a nondegenerate Floquet spectrum, the maximal divergence
\((1-a^2)^{1-p}\) occurs only when
\(\mu_1=\cdots=\mu_p\).  Terms containing fewer coincident indices have
lower-order poles and vanish after multiplication by
\(\epsilon^{p-1}=(1-a^2)^{p-1}\).  Therefore
\begin{align}
  &\lim_{\epsilon\downarrow0}
  \epsilon^{p-1}
  \int_0^{2\pi}\frac{\dd\phi}{2\pi}
  [G^+_{nn}(a,\phi)]^{p-1}
  G^-_{nn}(a,\phi)
  \nonumber\\
  &\qquad=
  \sum_{\mu=1}^{N}|\psi_\mu(n)|^{2p}.
  \label{eq:equal_state_projection}
\end{align}
Dividing by \(N\) reproduces the local moment in
Eq.~\eqref{eq:Mp}:
\begin{equation}
  M_p(n)
  =
  \frac1N
  \lim_{\epsilon\downarrow0}
  \epsilon^{p-1}
  \int_0^{2\pi}\frac{\dd\phi}{2\pi}
  [G^+_{nn}(a,\phi)]^{p-1}
  G^-_{nn}(a,\phi).
  \label{eq:Mp_resolvent}
\end{equation}
Indeed, the term in which all resolvents carry the same eigenstate contains
\begin{equation}
  \int_0^{2\pi}\frac{\dd\phi}{2\pi}
  \frac{1}{[1-a\e^{\iu\phi}]^{p-1}}
  \frac{1}{1-a\e^{-\iu\phi}}
  =
  (1-a^2)^{1-p},
  \label{eq:same_state_pole}
\end{equation}
whereas nondegenerate unequal-state terms are less singular.  Equation
\eqref{eq:Mp_resolvent} therefore fixes the normalization of the local
source insertion before any field-theory approximation.

\subsection{Master source functional}
\label{subsec:master_generator}

Introduce the Bose--Fermi source matrices
\begin{equation}
  \widehat e_+
  =
  \begin{pmatrix}a&0\\0&c\end{pmatrix}_{\BF},
  \qquad
  \widehat e_-
  =
  \begin{pmatrix}b&0\\0&d\end{pmatrix}_{\BF},
  \label{eq:ehat}
\end{equation}
and local bosonic rank-one sources
\begin{equation}
  J_\pm
  =
  \sum_n j_{\pm,n}
  \ket n\bra n\otimes P_B,
  \qquad
  P_B=\begin{pmatrix}1&0\\0&0\end{pmatrix}_{\BF}.
  \label{eq:Jpm}
\end{equation}
For each retarded or advanced sector, we introduce an \(N\)-component
supervector \(\psi_\pm\) in \(\QD\otimes\BF\).  Its component at momentum site \(n\) is
\begin{equation}
  \psi_{\pm,n}
  \equiv
  \begin{pmatrix}
    \psi_{\pm,B,n}\\
    \psi_{\pm,F,n}
  \end{pmatrix}_{\BF},
  \qquad n=1,\ldots,N,
  \label{eq:psi_components}
\end{equation}
where the \(B\) entry is an ordinary complex commuting variable and the
\(F\) entry is a Grassmann variable.  The corresponding \(\bar\psi_\pm\)
are row supervectors; the bosonic variables are integrated on the standard
convergent complex contours, while the barred and unbarred fermionic
variables are independent Grassmann variables.  Thus the sign \(\pm\) labels
\(\AR\), the index \(B/F\) labels \(\BF\), and \(n\) labels the quantum
Hilbert space \(\QD\).  The measure \(\dd(\psi,\bar\psi)\) below is the
product of the ordinary bosonic and Berezin fermionic measures over these
three spaces.  The capital supervector \(\Psi\) introduced after the
color--flavor transformation simply stacks \(\psi_+\) and \(\psi_-\) in
\(\AR\) space and introduces no additional degree of freedom.

The master functional is the normalized Gaussian superintegral
\begin{align}
  \cZ[\bm j;a,b,c,d]
  &=
  \int_0^{2\pi}\frac{\dd\phi}{2\pi}
  \int\dd(\psi,\bar\psi)
  \nonumber\\
  &\quad\times
  \exp\bigl[-\bar\psi_+A_+(\phi)\psi_+\bigr]
  \nonumber\\
  &\quad\times
  \exp\bigl[-\bar\psi_-A_-(\phi)\psi_-\bigr],
  \label{eq:masterZ}
\end{align}
with
\begin{align}
  A_+(\phi)
  &=1-\e^{\iu\phi}U_\chi\widehat e_+-J_+,
  \nonumber\\
  A_-(\phi)
  &=1-\e^{-\iu\phi}U_\chi^\dagger\widehat e_--J_-.
  \label{eq:Apm}
\end{align}
At \(\bm j=0\), define
\begin{align}
  D_+(z,\phi)&\equiv\det(1-z\e^{\iu\phi}U_\chi),\\
  D_-(z,\phi)&\equiv\det(1-z\e^{-\iu\phi}U_\chi^\dagger).
\end{align}
Gaussian integration then gives
\begin{equation}
  \cZ[0;a,b,c,d]
  =
  \int_0^{2\pi}\frac{\dd\phi}{2\pi}
  \frac{D_+(c,\phi)D_-(d,\phi)}
       {D_+(a,\phi)D_-(b,\phi)}.
  \label{eq:det_generator}
\end{equation}
At the supersymmetric point \(a=c\), \(b=d\), and \(\bm j=0\), one has
\(\cZ=1\).

For the local moments, define
\begin{equation}
  \mathcal D_n^{(p)}
  \equiv
  \frac1{(p-1)!}
  \frac{\partial^{p-1}}{\partial j_{+,n}^{p-1}}
  \frac{\partial}{\partial j_{-,n}}.
  \label{eq:Dpn}
\end{equation}
Then
\begin{equation}
  \av{I_p}
  =
  \frac1N
  \lim_{\epsilon\downarrow0}
  \epsilon^{p-1}
  \sum_n
  \left.
  \mathcal D_n^{(p)}
  \cZ[\bm j;a,b,a,b]
  \right|_{\bm j=0,\ a=b=\e^{-\gamma}}.
  \label{eq:Ip_source}
\end{equation}
We now derive Eq.~\eqref{eq:Ip_source} explicitly from the Gaussian source
functional.  At fixed \(\phi\), let \(A_{+,B}^{(0)}\) and
\(A_{-,B}^{(0)}\) denote the source-free bosonic blocks of
Eq.~\eqref{eq:Apm}.  The source at site \(n\) changes these blocks to
\(A_{\pm,B}^{(0)}-j_{\pm,n}\ket n\bra n\).  Factoring out the source-free
matrix gives
\begin{equation}
  A-j\ket n\bra n
  =A\left[1-jA^{-1}\ket n\bra n\right].
  \label{eq:rank_one_factorization}
\end{equation}
For a rank-one operator, the matrix determinant lemma gives
\begin{align}
  \det\left[1-jA^{-1}\ket n\bra n\right]
  &=1-j\bra nA^{-1}\ket n,
  \nonumber\\
  \det(A-j\ket n\bra n)^{-1}
  &=\det(A)^{-1}
  \left[1-j(A^{-1})_{nn}\right]^{-1}.
  \label{eq:rank_one_source_identity}
\end{align}
At the supersymmetric point, the source-free bosonic determinant is cancelled
by its fermionic partner.  The remaining source-dependent factors are
therefore
\begin{align}
  \mathcal B_+(j_{+,n})
  &=\left[1-j_{+,n}G^+_{nn}(a,\phi)\right]^{-1},
  \nonumber\\
  \mathcal B_-(j_{-,n})
  &=\left[1-j_{-,n}G^-_{nn}(b,\phi)\right]^{-1}.
  \label{eq:local_source_factors}
\end{align}
Expanding them as geometric series shows directly that
\begin{align}
  \left.
  \frac1{(p-1)!}
  \frac{\partial^{p-1}\mathcal B_+}
       {\partial j_{+,n}^{p-1}}
  \right|_{j_{+,n}=0}
  &=[G^+_{nn}(a,\phi)]^{p-1},
  \label{eq:retarded_source_derivative}\\
  \left.
  \frac{\partial\mathcal B_-}{\partial j_{-,n}}
  \right|_{j_{-,n}=0}
  &=G^-_{nn}(b,\phi).
  \label{eq:advanced_source_derivative}
\end{align}
The retarded and advanced sources act in different blocks, so their
 derivatives multiply.  Substitution into the master integral gives the
exact identity
\begin{align}
  &\left.
  \mathcal D_n^{(p)}
  \cZ[\bm j;a,b,a,b]
  \right|_{\bm j=0}
  \nonumber\\
  &\quad=
  \int_0^{2\pi}\frac{\dd\phi}{2\pi}
  [G^+_{nn}(a,\phi)]^{p-1}
  G^-_{nn}(b,\phi).
  \label{eq:source_to_resolvents}
\end{align}
Setting \(a=b=\e^{-\gamma}\), multiplying by
\(\epsilon^{p-1}\), and applying the equal-state projection
Eq.~\eqref{eq:equal_state_projection} gives, step by step,
\begin{align}
  &\frac1N
  \lim_{\epsilon\downarrow0}
  \epsilon^{p-1}
  \sum_n
  \left.
  \mathcal D_n^{(p)}\cZ
  \right|_{\bm j=0}
  \nonumber\\
  &\quad=
  \frac1N\sum_n\sum_{\mu=1}^{N}
  |\psi_\mu(n)|^{2p}
  \nonumber\\
  &\quad=
  \sum_nM_p(n)
  =\frac1N\sum_{\mu,n}|\psi_\mu(n)|^{2p}
  =\av{I_p}.
  \label{eq:source_equals_IPR}
\end{align}
Thus Eq.~\eqref{eq:Ip_source} follows from the rank-one source derivatives,
the exact \(\phi\) integral, and the equal-state pole; it is not an additional
definition of the IPR.

The same functional generates the quasienergy two-level statistics.  We first
fix the convention, because the connected correlator, the off-diagonal pair
correlator, and the cluster function have different Poisson limits.  The mean
eigenphase spacing is
\begin{equation}
  \Delta\equiv\frac{2\pi}{N}.
  \label{eq:Delta_definition}
\end{equation}
The two resolvents in the generating function are shifted by \(\pm e/N\).
Their physical eigenphase separation is therefore
\begin{equation}
  \Omega\equiv\frac{2e}{N},
  \qquad
  s\equiv\frac{\Omega}{\Delta}=\frac{e}{\pi}.
  \label{eq:unfolded_s_definition}
\end{equation}
We introduce the unfolded density
\begin{equation}
  \varrho(x)
  \equiv
  \Delta\,
  \rho(\phi_0+\Delta x),
  \qquad
  \av{\varrho(x)}=1,
  \label{eq:unfolded_density}
\end{equation}
For the density observables in this subsection,
\(\av{\cdots}\) denotes the spectral center-phase average: the origin
\(\phi_0\) of the unfolded quasienergy spectrum is averaged uniformly over one
Floquet zone.  It is not a disorder average or an average over different
Floquet operators.  The off-diagonal pair correlation is
\begin{equation}
  R_2(s)
  \equiv
  \av{
    \sum_{\mu\neq\nu}
    \delta\!\left(x+\frac{s}{2}-x_\mu\right)
    \delta\!\left(x-\frac{s}{2}-x_\nu\right)
  }.
  \label{eq:R2_offdiag_definition}
\end{equation}
It is normalized so that
\begin{equation}
  R_2^{\rm Poisson}(s)=1.
  \label{eq:R2_Poisson_definition}
\end{equation}
We also define the cluster function and the connected density covariance,
respectively,
\begin{align}
  Y_2(s)
  &\equiv
  1-R_2(s),
  \label{eq:Y2_definition}\\
  \mathcal K_2(s)
  &\equiv
  \av{\delta\varrho(x+s/2)\delta\varrho(x-s/2)}
  \nonumber\\
  &=
  \delta(s)-Y_2(s)
  =
  \delta(s)+R_2(s)-1.
  \label{eq:K2_definition}
\end{align}
Thus Poisson statistics means \(Y_2=0\), \(\mathcal K_2=\delta\), and
\(R_2=1\).  CUE statistics instead has a nonzero negative connected part.

The global determinant source is
\begin{equation}
  C(e)
  =
  \left.
  \frac{2ab}{N^2}
  \partial_c\partial_d
  \cZ[0;a,b,c,d]
  \right|_{a=b=c=d=\exp(\iu e/N-0^+)}.
  \label{eq:Csource}
\end{equation}
Using
\begin{equation}
  \Re\frac{\e^{\iu(x+\iu0)}}{1-\e^{\iu(x+\iu0)}}
  =-\frac12+\pi\delta_{2\pi}(x),
  \label{eq:distribution_identity}
\end{equation}
and the relation \(e=\pi s\), the source produces the connected object
\begin{equation}
  \mathcal K_2(s)=\Re C(\pi s).
  \label{eq:K2_from_C}
\end{equation}
For \(s\neq0\), where the self-correlation delta function is absent,
\begin{equation}
  R_2(s)=1+\Re C(\pi s).
  \label{eq:R2_from_C}
\end{equation}
Local derivatives of \(\cZ\) therefore generate \(I_p\), while global
determinant derivatives generate \(\mathcal K_2\), and hence \(R_2\).

\subsection{Color--flavor transformation}
\label{subsec:CFT}

The center-phase integral is performed exactly with the \(U(1)\)
color--flavor identity of Refs.~\cite{ZirnbauerCircular1996,
ZirnbauerCFT1998,ZirnbauerCFT2021}.  For arbitrary graded supervectors
\(\Psi_1,\Psi_2,\Psi'_1,\Psi'_2\), the identity reads
\begin{align}
  &\int_0^{2\pi}\frac{\dd\phi}{2\pi}
  \exp\left[
    \e^{\iu\phi}\Psi_1^{T}\Psi'_2
    +
    \e^{-\iu\phi}\Psi_2^{T}\Psi'_1
  \right]
  \nonumber\\
  &\quad=
  \int\dd(Z,\widetilde Z)\,
  \sdet(1-Z\widetilde Z)
  \nonumber\\
  &\qquad\times
  \exp\left[
    \Psi_1^{T}\widetilde Z\Psi'_1
    +
    \Psi_2^{T}Z\Psi'_2
  \right].
  \label{eq:color_flavor_identity}
\end{align}
Here \(Z\) and \(\widetilde Z\) act in
\(\BF\otimes\QD\); the transpose is the graded transpose appropriate to
the supervectors.  In the boson--boson and fermion--fermion sectors the
convergence contours are
\(\widetilde Z_{BB}=Z_{BB}^{\dagger}\) and
\(\widetilde Z_{FF}=-Z_{FF}^{\dagger}\), with the bosonic eigenvalues
restricted inside the unit disk.  Applying
Eq.~\eqref{eq:color_flavor_identity} to the two phase-dependent
bilinears in Eq.~\eqref{eq:masterZ} replaces the center-phase integral by
the \(Z,\widetilde Z\) integral.  At zero local source one obtains
\begin{equation}
  \cZ
  =
  \int\dd(Z,\widetilde Z)\,
  \e^{-\cS[Z,\widetilde Z]},
  \label{eq:Zsuper}
\end{equation}
where
\begin{align}
  \cS[Z,\widetilde Z]
  &=-\str\ln(1-Z\widetilde Z)
  \nonumber\\
  &\quad+
  \str\ln\left[
    1-U_\chi\widetilde ZU_\chi^\dagger
    \widehat e_-Z\widehat e_+
  \right].
  \label{eq:exactaction}
\end{align}
The fields act in \(\BF\otimes\QD\); the advanced--retarded structure is
encoded by the placement of \(Z\) and \(\widetilde Z\).  We now retain the
local sources and display the remaining Gaussian integral explicitly.
Introduce the combined retarded--advanced supervector
\begin{equation}
  \Psi
  \equiv
  \begin{pmatrix}
    \psi_+\\
    \psi_-
  \end{pmatrix}_{\AR},
  \qquad
  \bar\Psi
  \equiv
  \begin{pmatrix}
    \bar\psi_+&\bar\psi_-
  \end{pmatrix}_{\AR},
  \label{eq:Psi_AR_definition}
\end{equation}
and the block operator
\begin{equation}
  \mathcal M[\bm j]
  =
  \begin{pmatrix}
    1-J_+
    &-U_\chi\widetilde ZU_\chi^\dagger\\[1mm]
    -\widehat e_-Z\widehat e_+
    &1-J_-
  \end{pmatrix}_{\AR},
  \label{eq:Msource}
\end{equation}
An alternative source bookkeeping, used in
Ref.~\cite{LiaoGalitski2022}, keeps \(J_\pm\) as full matrices in the
Hilbert space and combines them multiplicatively with the Bose--Fermi
spectral-parameter blocks of the generating function.  After the
color--flavor transformation this is equivalent to working with
source-dressed parameter matrices rather than with the additive local terms
in \(\mathcal M[\bm j]\).  That form is convenient when all four
Hilbert-space indices of an eigenstate correlator are to be retained.  The
diagonal bosonic sources in Eq.~\eqref{eq:Jpm} are the specialization
needed here: each local derivative projects directly onto an intensity at a
fixed momentum site and keeps the equal-state pole normalization of
Eq.~\eqref{eq:Ip_source} explicit.  We therefore retain the present
convention.
The color--flavor-transformed generating functional is then
\begin{align}
  \cZ[\bm j;a,b,c,d]
  &=
  \int\dd(Z,\widetilde Z)\,
  \sdet(1-Z\widetilde Z)
  \nonumber\\
  &\quad\times
  \int\dd(\Psi,\bar\Psi)\,
  \exp\left[
    -\bar\Psi\mathcal M[\bm j]\Psi
  \right].
  \label{eq:Z_with_M_integral}
\end{align}
The remaining super-Gaussian integral is
\begin{equation}
  \int\dd(\Psi,\bar\Psi)\,
  \e^{-\bar\Psi\mathcal M\Psi}
  =
  \Sdet\nolimits^{-1}\mathcal M,
  \label{eq:Gaussian_Sdet_M}
\end{equation}
so that
\begin{align}
  \cZ[\bm j;a,b,c,d]
  &=
  \int\dd(Z,\widetilde Z)\,
  \sdet(1-Z\widetilde Z)
  \Sdet\nolimits^{-1}\mathcal M[\bm j].
  \label{eq:Z_exact_with_sources}
\end{align}
At \(\bm j=0\), the block-superdeterminant identity gives
\begin{align}
  \Sdet\mathcal M[0]
  &=
  \sdet\left[
    1-U_\chi\widetilde ZU_\chi^\dagger
    \widehat e_-Z\widehat e_+
  \right],
  \label{eq:M_Schur_complement}
\end{align}
where cyclicity of \(\str\ln\) has been used to place the factors in the
same order as Eq.~\eqref{eq:exactaction}.  Combining
Eqs.~\eqref{eq:Z_exact_with_sources} and
\eqref{eq:M_Schur_complement} reproduces the exact action
Eq.~\eqref{eq:exactaction}.
For the global source sector define
\begin{align}
  X&\equiv U_\chi\widetilde ZU_\chi^\dagger\widehat e_-Z\widehat e_+,
  \qquad \mathcal G\equiv(1-X)^{-1},
  \label{eq:XG}\\
  X_c&\equiv U_\chi\widetilde ZU_\chi^\dagger\widehat e_-ZP_F,
  \nonumber\\
  X_d&\equiv U_\chi\widetilde ZU_\chi^\dagger P_FZ\widehat e_+,
  \nonumber\\
  X_{cd}&\equiv U_\chi\widetilde ZU_\chi^\dagger P_FZP_F.
  \label{eq:Xcd}
\end{align}
Here \(P_F=\operatorname{diag}(0,1)_{\BF}\).  We define
\(\mathcal O_C[Z,\widetilde Z]\) as the normalized mixed derivative of the
field-theory weight with respect to the two global fermionic source
parameters \(c\) and \(d\).  It is therefore the exact \(Z,\widetilde Z\)
operator whose expectation value produces the two-level generating
function \(C(e)\).  Equivalently,
\begin{equation}
  \mathcal O_C
  =
  (\partial_c\cS)(\partial_d\cS)
  -
  \partial_c\partial_d\cS.
  \label{eq:OC_action_derivatives}
\end{equation}
Using
\(\partial_c\cS=-\str(\mathcal G X_c)\),
\(\partial_d\cS=-\str(\mathcal G X_d)\), and
\(\partial_d\mathcal G=\mathcal G X_d\mathcal G\), differentiation of the full weight gives
\begin{align}
  \mathcal O_C[Z,\widetilde Z]
  &\equiv
  \frac{\partial_c\partial_d\e^{-\cS}}{\e^{-\cS}}
  \nonumber\\
  &=
  \str(\mathcal G X_c)\str(\mathcal G X_d)
  \nonumber\\
  &\quad+
  \str(\mathcal G X_d\mathcal G X_c)+\str(\mathcal G X_{cd}).
  \label{eq:OC_exact}
\end{align}
The first term is the product of the two one-source vertices.  The second
comes from differentiating the resolvent \(\mathcal G\), and the third is the direct
mixed derivative of \(X\).  All partial-derivative identities entering
Eqs.~\eqref{eq:OC_action_derivatives} and \eqref{eq:OC_exact} are derived
explicitly in Appendix~\ref{app:correlator}.
Thus the action, the local IPR insertion, and the global two-level insertion
are generated before the slow-mode approximation from one exact finite-$N$
object.
For parallel notation, the local moment insertion can be displayed at the
same exact finite-\(N\) stage.  Since the factor
\(\sdet(1-Z\widetilde Z)\) is independent of the local sources, define
\begin{equation}
  \mathcal O_p(n;Z,\widetilde Z)
  \equiv
  \left.
  \frac{
    \mathcal D_n^{(p)}\Sdet^{-1}\mathcal M[\bm j]
  }{
    \Sdet^{-1}\mathcal M[0]
  }
  \right|_{\bm j=0}.
  \label{eq:Op_exact_Z}
\end{equation}
With this definition, Eq.~\eqref{eq:Ip_source} is the expectation value of
\(\mathcal O_p(n;Z,\widetilde Z)\) with the same local-source-free
\(Z,\widetilde Z\) weight that appears in Eq.~\eqref{eq:Zsuper}, evaluated
at the supersymmetric point \(c=a\), \(d=b\) and finally
\(a=b=\e^{-\gamma}\).  Thus \(\mathcal O_p\) and \(\mathcal O_C\) are both
fixed before the slow-mode projection; the former is generated by local
bosonic derivatives and the latter by the global fermionic derivatives.

\subsection{Slow mode and nonlinear action}
\label{subsec:slow}
The quadratic expansion used here has a specific and limited purpose: it is
the Hessian analysis that identifies which operator-space directions are
soft.  It is not yet the perturbative loop expansion, and it does not assume
that the final nonlinear \(Q\) field has a small amplitude.  The control
parameter at this stage is the separation of relaxation scales.  The
regulator \(\epsilon\), the external frequency \(|\omega|\), and the
long-wavelength density relaxation eigenvalues \(1-\widehat P(q)\) are
taken small compared with the finite decay rates of the operator modes that
are discarded.  In this regime the quadratic Hessian is sufficient to decide
which directions become massless in the infrared; higher powers of
\(Z,\widetilde Z\) describe interactions among those directions but do not
change their identification as the soft sector.  Once that sector has been
isolated, the class-A symmetry restores its full nonlinear completion.  This
is precisely the logic used for the kicked rotor: one first restricts to
lowest order in \(Z,\widetilde Z\) to identify the slowly varying diagonal
modes and then removes the small-field restriction by rewriting the theory in
terms of \(Q=T\Lambda T^{-1}\)
\cite{AltlandZirnbauerQKR1996,AltlandEtAl2015}.  The small parameter of the
later fluctuation/RG expansion is a different quantity, the running coupling
whose bare value is \(t_0\) in Eq.~\eqref{eq:t0_definition}.
At the supersymmetric point, we follow the slow-mode projection used for
quantum maps and kicked rotors~\cite{AltlandZirnbauerQKR1996,
TianAltland2010,AltlandEtAl2015}.  Expanding
Eq.~\eqref{eq:exactaction} to quadratic order gives, at
\(\epsilon=\omega=0\),
\begin{equation}
  \cS^{(2)}
  =
  \str\left(
    Z\widetilde Z
    -
    U_\chi\widetilde ZU_\chi^\dagger Z
  \right).
  \label{eq:S2_operator_space}
\end{equation}
Thus the masses of the quadratic modes are determined by the adjoint
Floquet superoperator
\begin{equation}
  \mathcal P_U(\mathsf A)
  \equiv
  U_\chi\mathsf A U_\chi^\dagger
  \label{eq:adjoint_Floquet}
\end{equation}
acting in operator space.  A slow field must be supported on eigenoperators
of \(\mathcal P_U\) whose eigenvalues approach unity.
Here and in the following operator-space argument, \(\mathsf A\) denotes a
generic Hilbert-space operator.  This notation is distinct from the
source-dependent matrix \(X\) defined in Eq.~\eqref{eq:XG}.
To identify those modes, define the projector onto momentum-diagonal
operators,
\begin{equation}
  \mathcal P_{\rm d}\mathsf A
  \equiv
  \sum_n
  \ket n\bra n\,\mathsf A\,\ket n\bra n.
  \label{eq:diagonal_projector}
\end{equation}
The free-rotation part of \(U_\chi\) multiplies an off-diagonal operator
\(\ket n\bra m\), \(n\neq m\), by the phase
\begin{equation}
  \exp\left[
    -\frac{\iu\hbareff}{2}
    \bigl(
      (n+\chi)^2-(m+\chi)^2
    \bigr)
  \right].
  \label{eq:offdiagonal_phase}
\end{equation}
The absence of a conserved off-diagonal coherence can be seen explicitly.
Write \(r=n-m\neq0\) and use the center coordinate
\(R=(n+m)/2\).  The phase in Eq.~\eqref{eq:offdiagonal_phase} is
\begin{equation}
  \vartheta_{R,r}
  =\hbareff r(R+\chi).
  \label{eq:coherence_phase_center}
\end{equation}
A translation of the center coordinate by one momentum site changes the
phase by the nonzero amount
\begin{equation}
  \vartheta_{R+1,r}-\vartheta_{R,r}=\hbareff r.
  \label{eq:coherence_phase_increment}
\end{equation}
The coarse-grained amplitude of such a coherence over \(L\) center positions
is the geometric sum
\begin{align}
  \mathfrak A_L(r)
  &\equiv
  \frac1L\sum_{R=0}^{L-1}\e^{-\iu\hbareff rR}
  \nonumber\\
  &=
  \frac{1-\e^{-\iu\hbareff rL}}
       {L[1-\e^{-\iu\hbareff r}]}.
  \label{eq:coherence_geometric_average}
\end{align}
For irrational \(\hbareff/(4\pi)\), the denominator in
Eq.~\eqref{eq:coherence_geometric_average} is nonzero for every integer
\(r\neq0\), while the numerator remains bounded.  Hence
\begin{equation}
  \lim_{L\to\infty}\mathfrak A_L(r)=0,
  \qquad r\neq0,
  \label{eq:coherence_average_zero}
\end{equation}
whereas \(\mathfrak A_L(0)=1\).  The generic flux excludes the residual
pairing degeneracies that would otherwise relate opposite momenta.  Thus an
off-diagonal operator has no phase-independent component protected by a
continuity equation, and no eigenvalue of the coarse-grained adjoint map is
pinned to unity in that sector.

The distinction can also be read directly from the action of the full
Floquet map.  For a diagonal operator
\begin{equation}
  \mathsf A_{\rm d}=\sum_m\mathsf a_m\ket m\bra m,
  \label{eq:diagonal_operator_expansion}
\end{equation}
Eq.~\eqref{eq:Unm} gives
\begin{align}
  [\mathcal P_U(\mathsf A_{\rm d})]_{nn'}
  &=
  \e^{-\iu\hbareff[(n+\chi)^2-(n'+\chi)^2]/2}
  \nonumber\\
  &\quad\times
  \sum_m u_{n-m}u_{n'-m}^{*}\mathsf a_m.
  \label{eq:adjoint_on_diagonal_operator}
\end{align}
For \(n\neq n'\), the prefactor is precisely the rapidly varying coherence
phase just averaged in Eqs.~\eqref{eq:coherence_phase_center}--
\eqref{eq:coherence_average_zero}.  For \(n=n'\), it cancels identically and
one obtains
\begin{align}
  \left[
    \mathcal P_{\rm d}
    \mathcal P_U(\mathsf A_{\rm d})
  \right]_{nn}
  &=
  \sum_m
  |u_{n-m}|^2\mathsf a_m
  \nonumber\\
  &=
  \sum_m
  P_{n-m}\mathsf a_m.
  \label{eq:projected_probability_operator}
\end{align}
Moreover,
\begin{align}
  \sum_n
  \left[
    \mathcal P_{\rm d}\mathcal P_U(\mathsf A_{\rm d})
  \right]_{nn}
  &=
  \sum_m\mathsf a_m\sum_rP_r
  =\sum_m\mathsf a_m,
  \label{eq:probability_conservation_projected}
\end{align}
so the uniform diagonal density is protected at eigenvalue one by
probability conservation.  No analogous identity exists for
\(n\neq n'\).  Off-diagonal coherences therefore acquire a finite decay
rate under coarse graining, while diagonal momentum probabilities contain
the hydrodynamic mode.

Projecting out the gapped off-diagonal operator modes gives
\begin{equation}
  Z_{nm}\simeq\delta_{nm}Z_n,
  \qquad
  \widetilde Z_{nm}\simeq\delta_{nm}\widetilde Z_n.
  \label{eq:slow}
\end{equation}
This is an infrared hydrodynamic projection, not a statement that the
Hilbert-space off-diagonal modes are irrelevant to every eigenfunction
observable.  A general four-index eigenstate correlator is generated by full
matrix sources and can receive contributions from Hilbert-space off-diagonal
fluctuation sectors; this structure is kept explicitly in the field-theory
calculation of Ref.~\cite{LiaoGalitski2022}.  Equation~\eqref{eq:slow} is
therefore used here only after specifying the local IPR and global spectral
sources for which the long-distance hydrodynamic sector is sufficient.
Basis-resolved correlations at finite scales can retain additional massive-
mode contributions and would require extending the projection accordingly.
Using Eq.~\eqref{eq:Unm},
\begin{equation}
  \str(U_\chi\widetilde ZU_\chi^\dagger Z)
  =
  \sum_{n,m}P_{n-m}\str(Z_n\widetilde Z_m),
  \label{eq:slowcontraction}
\end{equation}
where
\begin{equation}
  P_r\equiv|u_r|^2,
  \qquad
  \sum_rP_r=1.
  \label{eq:Pr}
\end{equation}
Because the inversion-breaking harmonic makes the directed probabilities
slightly asymmetric, we also define
\begin{equation}
  P_r^{(s)}\equiv\frac{P_r+P_{-r}}2.
  \label{eq:Psym}
\end{equation}
The quadratic action is
\begin{align}
  \cS^{(2)}
  &=
  \int_{-\pi}^{\pi}\frac{\dd q}{2\pi}\,
  \str[Z(q)\widetilde Z(-q)]
  \nonumber\\
  &\quad\times
  [\epsilon-\iu\omega+1-\widehat P(q)],
  \label{eq:S2q}
\end{align}
with
\begin{align}
  \widehat P(q)
  &=\sum_rP_r\e^{-\iu qr}
  \nonumber\\
  &=
  \int_{-\pi}^{\pi}\frac{\dd x}{2\pi}
  \exp\{-\iu\kappa[V_{\beta,\eta}(x)-V_{\beta,\eta}(x+q)]\}.
  \label{eq:Phat}
\end{align}
We define the relaxation eigenvalue
\begin{equation}
  \lambda(q)\equiv1-\widehat P(q).
  \label{eq:lambdadef}
\end{equation}
Probability conservation implies \(\lambda(0)=0\).  

The fields \(Z_n\) and \(\widetilde Z_n\) are the coordinates produced
directly by the color--flavor transformation.  They describe small rotations
between the retarded and advanced saddle sectors, but they do not display the
nonlinear symmetry globally.  We therefore introduce the matrix
\begin{equation}
  T_n\equiv
  \begin{pmatrix}
    1&Z_n\\
    \widetilde Z_n&1
  \end{pmatrix}_{\AR},
  \label{eq:T_from_Z}
\end{equation}
The retarded--advanced grading matrix is
\begin{equation}
  \Lambda
  \equiv
  \begin{pmatrix}
    1&0\\
    0&-1
  \end{pmatrix}_{\AR},
  \label{eq:Lambda_AR_definition}
\end{equation}
with the identity in Bose--Fermi space understood.  We then use \(T_n\) to
define the nonlinear field
\begin{equation}
  Q_n\equiv T_n\Lambda T_n^{-1}.
  \label{eq:Qdef}
\end{equation}
This definition implies
\begin{equation}
  Q_n^2
  =
  T_n\Lambda^2T_n^{-1}
  =
  1,
  \label{eq:Qconstraint}
\end{equation}
because \(\Lambda^2=1\).  Thus \(Q_n\) lies on the class-A target space
\(U(1,1|2)/[U(1|1)\times U(1|1)]\), in the standard supersymmetric
classification~\cite{Efetov1983,EfetovBook,AltlandZirnbauer1997}.

To match the microscopic color--flavor coordinates to the standard nonlinear variable at the order required to identify the kinetic kernel, we expand the inverse of
\(T_n\) for small \(Z_n,\widetilde Z_n\).  This gives
\begin{align}
  Q_n^{RA}
  &=
  -2Z_n+O(Z^3),
  \nonumber\\
  Q_n^{AR}
  &=
  2\widetilde Z_n+O(\widetilde Z^3).
  \label{eq:Qblocks_from_Z}
\end{align}
Equations~\eqref{eq:Qblocks_from_Z} and \eqref{eq:S2q} establish the
matching of the microscopic \(Z,\widetilde Z\) description and the nonlinear
\(Q\)-field description to quadratic order about the uniform saddle.  This
comparison by itself does not derive all higher-order vertices.  Their
nonlinear completion is fixed by the constraint \(Q^2=1\), the class-A
invariance of the slow manifold, and the standard kicked-rotor sigma-model
construction~\cite{AltlandZirnbauerQKR1996,TianAltland2010,AltlandEtAl2015}.
We use \(Q\) from this point onward for that symmetry-completed slow theory.
Its nonlocal kernel is derived below from the singular kick.  The resulting
lattice action is
\begin{align}
  \cS[Q]
  &=
  \frac{\epsilon-\iu\omega}{4}
  \sum_n\Str(\Lambda Q_n)
  \nonumber\\
  &\quad-
  \frac1{16}
  \sum_{n,m}P_{n-m}^{(s)}
  \Str(Q_n-Q_m)^2.
  \label{eq:latticesigma}
\end{align}
Only the symmetrized kernel appears here because
\(\Str(Q_n-Q_m)^2\) is invariant under \(n\leftrightarrow m\); the
antisymmetric part of \(P_{n-m}\) cancels exactly in the double sum.
The exact local and global source operators now descend through the same
slow-mode projection.  We denote the slow-manifold image of
Eq.~\eqref{eq:Op_exact_Z} by
\begin{equation}
  \mathcal O_p(n;Q)
  \equiv
  \left.
  \mathcal O_p(n;Z,\widetilde Z)
  \right|_{\rm slow\ manifold},
  \label{eq:OpQ}
\end{equation}
and analogously write \(\mathcal O_C[Q]\) for the slow-manifold image of
Eq.~\eqref{eq:OC_exact}.
Then Eq.~\eqref{eq:Ip_source} becomes
\begin{equation}
  \av{I_p}
  =
  \frac1N
  \lim_{\epsilon\downarrow0}
  \epsilon^{p-1}
  \sum_n
  \int\mathcal DQ\,
  \mathcal O_p(n;Q)\e^{-\cS[Q]}.
  \label{eq:IpQ}
\end{equation}
The global source formula correspondingly gives
\begin{equation}
  C(e)
  =
  \frac{2ab}{N^2}
  \int\mathcal DQ\,
  \mathcal O_C[Q]\e^{-\cS[Q]}.
  \label{eq:CQ}
\end{equation}
According to Eqs.~\eqref{eq:K2_from_C} and \eqref{eq:R2_from_C}, this
single insertion gives \(\mathcal K_2(s)\) and \(R_2(s)\).  We have
therefore obtained one class-A field theory with a local IPR insertion and a
global two-level insertion.
\section{Fourier kernel and scale regimes}
\label{sec:fourier}

We use the following notation throughout this section.  We denote by
\(r=n-m\) the discrete momentum jump generated by one kick, by \(q\) the
wave number Fourier conjugate to \(r\), by \(P_r\) the directed probability of the jump \(r\), by
\(P_r^{(s)}\) its symmetrized part, and by \(\widehat P(q)\) the
corresponding characteristic function.
The microscopic action and source insertions are now reduced to a concrete Fourier-analysis problem:
determine the large-\(|r|\) behavior of \(u_r\), square it to obtain
\(P_r\), and convert the symmetric long-range tail into the small-\(|q|\)
relaxation eigenvalue.  This order of operations is essential for singular unit-modulus phases and
is the microscopic step that distinguishes the present rotor from assuming
a random long-range hopping model.  The resulting nonlocal sigma-model
structure can then be compared with Refs.~\cite{FyodorovMirlin1991,
MirlinEtAl1996,MirlinReview2000}.  This section performs those steps and
identifies the
scale window in which a single fractional exponent is meaningful.

\subsection{From a probability tail to a fractional action}

We now connect the microscopic probability \(P_r\) to the coefficient that
appears in the continuum action.  Suppose that the symmetrized probability of a long
momentum jump has the algebraic tail
\begin{equation}
  P_r^{(s)}
  =
  \frac{A}{|r|^{1+\sigma}}
  [1+o(1)],
  \qquad
  0<\sigma<2.
  \label{eq:tail}
\end{equation}
Here \(A>0\) is the tail amplitude and \(\sigma\) is introduced as the
fractional transport exponent.  The range \(0<\sigma<2\) is singled out
because the second moment of \(P_r\) diverges and ordinary diffusion cannot
be obtained by a Taylor expansion in \(q\).

The odd harmonic in Eq.~\eqref{eq:V} makes the microscopic directed
probabilities \(P_r\) and \(P_{-r}\) unequal at finite \(r\).  The singular
large-distance contribution derived below is nevertheless symmetric to
leading order: the smooth odd factor changes only subleading powers.  The
leading nonlocal relaxation kernel is therefore the real, even part
\begin{equation}
  1-\Re\widehat P(q)
  =
  2\sum_{r=1}^{\infty}
  P_r^{(s)}[1-\cos(qr)].
  \label{eq:oneMinusPexact}
\end{equation}
The antisymmetric part contains only subleading corrections to the fractional
infrared term.  In particular there is no uniform drift: for the periodic
unit-modulus kick multiplier \(g(x)=\exp[-\iu\kappa V_{\beta,\eta}(x)]\),
Parseval's identity gives
\begin{equation}
  \sum_r rP_r
  =\int_{-\pi}^{\pi}\frac{\dd x}{2\pi}\,
  g^*(x)\,\iu\partial_x g(x)
  =\frac{\kappa}{2\pi}\int_{-\pi}^{\pi}\dd x\,V'_{\beta,\eta}(x)=0,
  \label{eq:no_drift}
\end{equation}
with the rounded-core regulator understood when needed.  From this point on,
when writing the continuum relaxation law we use \(\widehat P(q)\) for this
symmetrized infrared kernel; the discarded odd part is a subleading
microscopic correction.

At small \(|q|\), the dominant values in the algebraic part of the sum have
\(r\sim |q|^{-1}\).  We may therefore insert Eq.~\eqref{eq:tail}, replace
the sum by an integral, and set \(x=|q|r\):
\begin{align}
  1-\widehat P(q)
  &\simeq
  2A\int_0^\infty
  \frac{1-\cos(qr)}{r^{1+\sigma}}\dd r
  \nonumber\\
  &=
  2A|q|^\sigma
  \int_0^\infty
  \frac{1-\cos x}{x^{1+\sigma}}\dd x .
  \label{eq:tailsum}
\end{align}
The remaining dimensionless integral is
\begin{equation}
  \int_0^\infty
  \frac{1-\cos x}{x^{1+\sigma}}\dd x
  =
  \frac{\pi}{
    2\Gamma(1+\sigma)
    \sin(\pi\sigma/2)
  }.
  \label{eq:cosint}
\end{equation}
It follows that
\begin{align}
  1-\widehat P(q)
  &=
  D_\sigma|q|^\sigma
  +o(|q|^\sigma),
  \label{eq:fractionalkernel}\\
  D_\sigma
  &\equiv
  \frac{\pi A}{
    \Gamma(1+\sigma)
    \sin(\pi\sigma/2)
  }.
  \label{eq:Dsigma}
\end{align}
We call \(D_\sigma\) the generalized diffusion constant.  It converts the
microscopic tail amplitude \(A\) into the relaxation rate of a mode with wave
number \(q\).  In continuum notation,
\(|q|^\sigma\) is the Fourier symbol of the fractional Laplacian
\(( -\partial_n^2)^{\sigma/2}\).  Consequently, after \(t\) kicks,
\begin{equation}
  [\widehat P(q)]^t
  \simeq
  \exp[-D_\sigma|q|^\sigma t],
  \label{eq:Ppower}
\end{equation}
which is the fractional-diffusion law that will reappear in the sigma-model
propagator and in the wave-packet dynamics.
At \(\sigma=2\), a tail \(P_r\simeq A/r^3\) gives
\begin{equation}
  1-\widehat P(q)
  \simeq
  A q^2\ln\frac1{|q|}.
  \label{eq:sigma2}
\end{equation}
If \(\sum_rr^2P_r\) is finite, the leading term is instead
\begin{equation}
  1-\widehat P(q)
  =
  D_2q^2+o(q^2),
  \qquad
  D_2=\frac12\sum_rr^2P_r .
  \label{eq:D2}
\end{equation}

\subsection{Positive singularity exponent}

The positive-\(\beta\) problem contains two possible Fourier mechanisms:
a strict endpoint asymptotic and, at large \(\kappa\), a semiclassical saddle.
We compute both because they lead to different scale windows.  We first consider \(0<\beta<1\).  In this range the kick multiplier is
continuous at the singular point but has a nonanalytic cusp.  Long Fourier
wavelengths therefore receive an endpoint contribution that can be obtained
by expanding the multiplier locally.  Near \(x=0\), the full class-A kick gives
\begin{align}
  \e^{-\iu\kappa V_{\beta,\eta}(x)}
  &=
  \e^{\iu\kappa/\beta}
  \exp\left[
    -\frac{\iu\kappa}{\beta}|x|^\beta
    -\iu\kappa\eta x
    +O(x^2,|x|^{\beta+2})
  \right]
  \nonumber\\
  &=
  \e^{\iu\kappa/\beta}
  \left[
    1-\frac{\iu\kappa}{\beta}|x|^\beta
    -\iu\kappa\eta x
    +O(|x|^{2\beta},|x|^{\beta+1},x^2)
  \right].
  \label{eq:positiveexp}
\end{align}
For \(0<\beta<1\), the first nonanalytic term is still
\(|x|^\beta\); the inversion-breaking term is analytic and contributes only
faster-decaying Fourier components.  Therefore the leading singular
coefficient is unchanged by \(\eta\).  The constant and purely smooth terms
do not contribute to the algebraic large-\(|r|\) tail.  Using
\begin{equation}
  \int_0^\infty
  x^\beta\cos(rx)\dd x
  =
  -\Gamma(1+\beta)
  \sin\frac{\pi\beta}{2}\,
  r^{-1-\beta},
  \label{eq:powercos}
\end{equation}
under Abel continuation, one finds
\begin{align}
  u_r^{\mathrm{end}}
  &\simeq
  \e^{\iu\kappa/\beta}
  C_\beta(\kappa)
  |r|^{-1-\beta},
  \label{eq:uend}\\
  C_\beta(\kappa)
  &=
  \frac{
    \iu\kappa\Gamma(\beta)
    \sin(\pi\beta/2)
  }{\pi}.
  \label{eq:Cbeta}
\end{align}
Squaring the amplitude removes the unimportant global phase and gives the
symmetrized one-kick probability tail
\begin{equation}
  P_r^{(s),\mathrm{end}}
  \simeq
  A_\beta|r|^{-2-2\beta},
  \label{eq:Pend}
\end{equation}
with
\begin{equation}
  A_\beta
  =
  \left[
    \frac{
      \kappa\Gamma(\beta)
      \sin(\pi\beta/2)
    }{\pi}
  \right]^2 .
  \label{eq:Abeta}
\end{equation}
The Abel--Mellin derivation and the all-orders endpoint series are given
in Appendix~\ref{app:asymptotics}.  The endpoint expansion probes
\(x\sim1/r\) and is asymptotic when
\begin{equation}
  r\gg r_Q,
  \qquad
  r_Q\sim\max(1,\kappa^{1/\beta}).
  \label{eq:rQ}
\end{equation}

The power in Eq.~\eqref{eq:Pend} has the form
\(P_r\sim |r|^{-1-\sigma}\).  Matching exponents gives
\(\sigma=1+2\beta\).  Therefore, for \(0<\beta<1/2\),
Eqs.~\eqref{eq:Pend} and \eqref{eq:fractionalkernel} give
\begin{equation}
  \sigma_{\mathrm{IR}}=1+2\beta,
  \qquad
  1<\sigma_{\mathrm{IR}}<2 .
  \label{eq:sigmaIR}
\end{equation}
At \(\beta=1/2\) the kernel is
\(q^2\ln(1/|q|)\), and for \(\beta>1/2\) the finite second moment produces a
local \(q^2\) action.

There is also a semiclassical stationary-phase window.  On the \(x>0\)
branch, the full class-A phase is
\begin{equation}
  \Phi_\eta(x)
  =
  rx-\frac{\kappa}{\beta}x^\beta-\kappa\eta x+O(x^3).
  \label{eq:phase}
\end{equation}
Thus the smooth harmonic only shifts the large Fourier index to
\(r_{\rm eff}=r-\kappa\eta\), and
\begin{equation}
  x_s
  =
  \left(
    \frac{\kappa}{r-\kappa\eta}
  \right)^{1/(1-\beta)}
  =
  \left(\frac{\kappa}{r}\right)^{1/(1-\beta)}
  \left[1+O\left(\frac{\kappa\eta}{r}\right)\right].
  \label{eq:xs}
\end{equation}
When \(\max(\kappa,|\kappa\eta|)\ll r\ll\kappa^{1/\beta}\), this point lies
inside the singular region but remains separated from the endpoint.
Stationary phase gives
\begin{equation}
  P_r^{(s),\mathrm{sp}}
  \simeq
  \frac{
    \kappa^{1/(1-\beta)}
  }{
    2\pi(1-\beta)
  }
  |r|^{-1-\sigma_{\mathrm{sc}}}
  \left[1+O\left(\frac{\kappa\eta}{r}\right)\right],
  \label{eq:Psp}
\end{equation}
where
\begin{equation}
  \sigma_{\mathrm{sc}}
  =
  \frac1{1-\beta}.
  \label{eq:sigmasc}
\end{equation}
Thus the strict fixed-\(\kappa\) infrared exponent
\(\sigma_{\mathrm{IR}}\) and the finite-scale semiclassical exponent
\(\sigma_{\mathrm{sc}}\) are different.  The limits
\(r\to\infty\) and \(\kappa\to\infty\) do not commute.  The detailed
stationary-phase calculation is given in Appendix~\ref{app:asymptotics}.
The conclusion is that the exponent observed numerically depends on whether
the sampled momenta lie beyond or inside the stationary-phase window.  The
next subsection treats the marginal logarithmic kick, where the leading
critical coefficient can be obtained directly.

\subsection{Logarithmic kick}

At \(\beta=0\), the full class-A kick multiplier is
\begin{equation}
  \e^{-\iu\kappa V_{0,\eta}(x)}
  =
  \left(
    2\left|\sin\frac{x}{2}\right|
  \right)^{-\iu\kappa}
  \e^{-\iu\kappa\eta\sin x}.
  \label{eq:logmult}
\end{equation}
The long-range Fourier coefficient follows directly from the singular
neighborhood of \(x=0\).  Since
\begin{equation}
  \left(
    2\left|\sin\frac{x}{2}\right|
  \right)^{-\iu\kappa}
  \e^{-\iu\kappa\eta\sin x}
  =
  |x|^{-\iu\kappa}[1-\iu\kappa\eta x+O(x^2)],
  \label{eq:loglocal}
\end{equation}
the odd harmonic affects only the subleading endpoint terms.  The leading
Fourier integral is
\begin{align}
  u_r
  &\simeq
  \frac1\pi\int_0^\infty x^{-\iu\kappa}\cos(rx)\dd x
  \nonumber\\
  &=
  \frac{\Gamma(1-\iu\kappa)}{\pi}
  \cos\left[\frac{\pi}{2}(1-\iu\kappa)\right]
  |r|^{-1+\iu\kappa}
  \nonumber\\
  &=
  \frac{\iu\Gamma(1-\iu\kappa)}{\pi}
  \sinh\frac{\pi\kappa}{2}\,
  |r|^{-1+\iu\kappa}.
  \label{eq:ulog}
\end{align}
Using
\begin{equation}
  |\Gamma(1+\iu y)|^2
  =\frac{\pi y}{\sinh\pi y},
  \label{eq:gammamod_main}
\end{equation}
we obtain directly for the full \(\eta\neq0\) model
\begin{equation}
  P_r^{(s)}
  \simeq
  \frac{\kappa}{2\pi}
  \tanh\frac{\pi\kappa}{2}\,
  \frac1{r^2}.
  \label{eq:Plog}
\end{equation}
Thus Eq.~\eqref{eq:tail} has \(\sigma=1\) and
\begin{equation}
  1-\Re\widehat P(q)
  =D_1(\kappa)|q|+o(|q|),
  \qquad
  D_1(\kappa)
  =\frac{\kappa}{2}\tanh\frac{\pi\kappa}{2}.
  \label{eq:D1}
\end{equation}
The logarithmic rotor therefore realizes the marginal \(|q|\) kernel with an
exact leading bare coefficient even after inversion symmetry is broken.  The
full finite-\(q\) kernel remains the exact microscopic integral
Eq.~\eqref{eq:Phat} and contains \(\eta\)-dependent ultraviolet corrections,
but these do not change the critical coupling \(t_0=D_1^{-1}\) or the leading
infrared RG.  The Mellin steps underlying Eqs.~\eqref{eq:ulog}--\eqref{eq:D1}
are collected in Appendix~\ref{app:asymptotics}.  We next turn to negative
\(\beta\), where the Fourier integral is controlled by rapid phase
oscillations instead of an ordinary endpoint expansion.

\subsection{Negative exponent}

For \(-1<\beta<0\), the potential itself diverges and the phase of the kick
oscillates without limit near the origin.  This changes the dominant Fourier
mechanism: the endpoint cannot be treated by a uniform expansion in the
singular part of \(\kappa V_{\beta,\eta}\).  Indeed, near \(x=0\),
\begin{equation}
  \left|\kappa V_{\beta,\eta}(x)\right|
  \simeq
  \frac{\kappa}{|\beta|\,|x|^{|\beta|}},
  \label{eq:negative_expansion_parameter}
\end{equation}
which diverges as \(x\to0\).  The formal Taylor series
\begin{equation}
  \e^{-\iu\kappa V_{\beta,\eta}(x)}
  =
  1-\iu\kappa V_{\beta,\eta}(x)
  +O\!\left(\kappa^2V_{\beta,\eta}(x)^2\right)
  \label{eq:negative_formal_expansion}
\end{equation}
therefore has no \(x\)-independent small parameter on any interval containing
the singular point.  In an endpoint estimate for the Fourier coefficient,
the scale sampled at momentum \(r\) is \(x\sim r^{-1}\), so the effective
expansion parameter is
\begin{equation}
  \zeta_r
  \equiv
  \frac{\kappa}{|\beta|}r^{|\beta|}.
  \label{eq:zeta_r}
\end{equation}
Even when \(\kappa\ll1\), \(\zeta_r\) grows with \(r\); hence the
linearized endpoint expansion can hold only over a finite preasymptotic
window and necessarily fails in the true large-\(r\) limit.  We now compute
the unrounded asymptotic explicitly by stationary phase.  For \(r>0\), the stationary branch lies
on \(x>0\), where \(s(x)\simeq x\), and
\begin{align}
  u_r
  &\simeq
  \frac{\e^{\iu\kappa/\beta}}{2\pi}
  \int_0^\infty\dd x\,
  \exp[\iu\Phi_{-,\eta}(x)],
  \label{eq:negative_ur_integral}\\
  \Phi_{-,\eta}(x)
  &\equiv
  rx-\frac{\kappa}{\beta}x^\beta-\kappa\eta x+O(x^3).
  \label{eq:negative_phase}
\end{align}
The stationary-point equation is
\begin{equation}
  \Phi_{-,\eta}'(x_s)
  =
  r-\kappa\eta-\kappa x_s^{\beta-1}+O(x_s^2)
  =0,
\end{equation}
so that
\begin{equation}
  x_s
  =
  \left(
    \frac{\kappa}{r-\kappa\eta}
  \right)^{1/(1-\beta)}
  =
  \left(\frac{\kappa}{r}\right)^{1/(1-\beta)}
  \left[1+O\left(\frac{\kappa\eta}{r}\right)\right].
  \label{eq:negative_xs}
\end{equation}
Its curvature is positive,
\begin{align}
  \Phi_{-,\eta}''(x_s)
  &=
  \kappa(1-\beta)x_s^{\beta-2}+O(x_s)
  \nonumber\\
  &=
  \frac{(1-\beta)(r-\kappa\eta)}{x_s}[1+o(1)].
  \label{eq:negative_curvature}
\end{align}
The negative-\(x\) branch is nonstationary for \(r>0\) and is
subleading.  Applying the stationary-phase formula to the positive branch
gives
\begin{align}
  u_r
  &\simeq
  \frac{
    \e^{
      \iu\kappa/\beta
      +\iu\Phi_{-,\eta}(x_s)
      +\iu\pi/4
    }
  }{
    \sqrt{2\pi(1-\beta)}
  }
  \nonumber\\
  &\quad\times
  \kappa^{1/[2(1-\beta)]}
  r^{-(2-\beta)/[2(1-\beta)]}.
  \label{eq:negative_u_stationary}
\end{align}
Symmetrizing the two directed large-$|r|$ branches after squaring Eq.~\eqref{eq:negative_u_stationary} yields
\begin{equation}
  P_r^{(s)}
  \simeq
  \frac{
    \kappa^{1/(1-\beta)}
  }{
    2\pi(1-\beta)
  }
  |r|^{-1-\sigma_-}
  \left[1+O\left(\frac{\kappa\eta}{|r|}\right)\right],
  \label{eq:Pnegative}
\end{equation}
with
\begin{equation}
  \sigma_-=\frac1{1-\beta}<1 .
  \label{eq:sigmanegative}
\end{equation}
At weak kick there is nevertheless a finite preasymptotic interval in
which \(\zeta_r\ll1\).  In that interval we may retain the first
nonconstant term in Eq.~\eqref{eq:negative_formal_expansion}.  Since near the endpoint
\(V_{\beta,\eta}(x)\simeq(|x|^\beta-1)/\beta+\eta x\), the smooth
odd term is subleading and the constant term contributes only
to \(u_0\), whereas for \(r\neq0\)
\begin{align}
  u_r^{\mathrm{pre}}
  &\simeq
  -\frac{\iu\kappa\,\e^{\iu\kappa/\beta}}{\pi\beta}
  \int_0^\infty x^\beta\cos(rx)\dd x
  \nonumber\\
  &=
  \e^{\iu\kappa/\beta}
  \frac{
    \iu\kappa\Gamma(\beta)
    \sin(\pi\beta/2)
  }{\pi}
  |r|^{-1-\beta}.
  \label{eq:negative_pre_u}
\end{align}
In the second line we used Eq.~\eqref{eq:powercos} by Abel continuation,
which is valid for \(-1<\beta<0\).  Squaring the amplitude gives
\begin{align}
  P_r^{(s),\mathrm{pre}}
  &\simeq
  A_\beta^{\mathrm{pre}}
  |r|^{-2-2\beta},
  \label{eq:Ppre}\\
  A_\beta^{\mathrm{pre}}
  &\equiv
  \left[
    \frac{
      \kappa\Gamma(\beta)
      \sin(\pi\beta/2)
    }{\pi}
  \right]^2.
  \label{eq:Apre_negative}
\end{align}
The condition for this linearization is \(\zeta_r\ll1\).  Using
Eq.~\eqref{eq:zeta_r}, its upper crossover scale is therefore
\begin{equation}
  r_{\mathrm{nl}}
  \sim
  \left(
    \frac{|\beta|}{\kappa}
  \right)^{1/|\beta|}.
  \label{eq:rnl}
\end{equation}
For \(r\gtrsim r_{\mathrm{nl}}\), all powers of the singular phase are comparable and the stationary-phase result
Eq.~\eqref{eq:Pnegative}, rather than the linearized tail, controls the
asymptotics.
For \(\beta\leq-1/2\), Eq.~\eqref{eq:Ppre} is not summable and cannot define
the infinite-system kernel.  This is an explicit example in which a finite
matrix can look PRBM-like over an intermediate interval while the true
asymptotic exponent is different.  The conclusion is not merely that the
state is extended: one must also distinguish the preasymptotic and true
infrared transport exponents.  We now collect the cutoff conditions deciding
which exponent can actually be observed.

\subsection{Core and finite-size restrictions}

The preceding asymptotics become physical predictions only if their
stationary or endpoint regions lie between the microscopic core and the
finite-system infrared scale.  This subsection states that window explicitly.
The endpoint asymptotic requires
\begin{equation}
  r_Q\ll r\ll\delta_x^{-1}.
  \label{eq:endwindow}
\end{equation}
The stationary point must remain outside the core,
\(x_s\gg\delta_x\), which gives
\begin{equation}
  r\ll
  r_{\mathrm{core}}^{\mathrm{sp}}
  =
  \kappa\delta_x^{\beta-1}.
  \label{eq:spcore}
\end{equation}
A fractional field theory exists only when the lower crossover scale is
parametrically smaller than
\[
  \min\left(
    N,\delta_x^{-1},
    r_{\mathrm{core}}^{\mathrm{sp}}
  \right).
\]
Otherwise the full lattice action in Eq.~\eqref{eq:latticesigma}, rather than
a single-power continuum approximation, must be used.  The conclusion of
the Fourier analysis is therefore a scale-dependent continuum kernel,
together with explicit tests of when that kernel is observable.  We now use
it to renormalize the action and to evaluate the source-generated
wave-function moments and their connected fluctuations.

\section{Class-A renormalization group and observables of the singular rotor}
\label{sec:predictions}
Two saddle configurations enter the calculations below.  The uniform, or
normal, saddle is \(Q_n=\Lambda\) at every momentum site.  It is spatially
homogeneous and is the reference configuration for the Gaussian propagator,
the RG, and the IPR expansion.  Spectral correlations also receive a
nonperturbative contribution from the nonstandard Andreev--Altshuler (AA)
saddles, obtained by a fermionic permutation that exchanges retarded and
advanced saddle eigenvalues.  These saddles provide the oscillatory
contribution complementary to the perturbative normal-saddle result
\cite{AndreevAltshuler1995,AltlandEtAl2015}.  Unless stated otherwise,
``saddle'' below refers to the uniform normal saddle; the AA saddles enter
only when the two-level spectral correlator is reconstructed.  For the
single retarded--advanced pair used in the present two-level generating
function, convenient representatives of the two saddle sectors are
\begin{align}
  (Q_{\rm N})_n^{BB}
  &=
  \Lambda,
  \qquad
  (Q_{\rm N})_n^{FF}
  =
  \Lambda,
  \label{eq:normal_saddle_explicit}\\
  (Q_{\rm AA})_n^{BB}
  &=
  \Lambda,
  \qquad
  (Q_{\rm AA})_n^{FF}
  =
  -\Lambda,
  \label{eq:AA_saddle_explicit}
\end{align}
for every momentum site \(n\).  Thus the AA saddle leaves the bosonic block on
the causal saddle and reverses the retarded--advanced signature only in the
fermionic block, which is precisely the fermionic permutation described
above~\cite{AndreevAltshuler1995,AltlandEtAl2015}.

The preceding sections produced the nonlinear field \(Q\), its microscopic
kernel, and the IPR and two-level source insertions.  To perform perturbation
theory we now expand \(Q\) about its uniform saddle.  The sequence is
\[
  (Z,\widetilde Z)
  \longrightarrow Q
  \longrightarrow W
  \longrightarrow (B,\widetilde B).
\]
These symbols do not denote four independent theories.  The pair
\(Z,\widetilde Z\) gives the color--flavor coordinates.  The field \(Q\) is
their nonlinear completion, \(W\) is a small tangent fluctuation of \(Q\),
and \(B,\widetilde B\) are its retarded--advanced blocks.

Assume that, in the scaling interval determined in
Sec.~\ref{sec:fourier},
\begin{equation}
  1-\widehat P(q)=D_\sigma|q|^\sigma.
  \label{eq:powerkernel}
\end{equation}
We introduce
\begin{equation}
  t_0\equiv D_\sigma^{-1},
  \label{eq:t0_definition}
\end{equation}
because powers of \(t_0\) count Gaussian contractions.  Substitution into
the lattice action gives
\begin{align}
  \cS[Q]
  &=
  -\frac1{8t_0}
  \int\frac{\dd q}{2\pi}
  |q|^\sigma\Str(Q_qQ_{-q})
  \nonumber\\
  &\quad+
  \frac{\epsilon-\iu\omega}{4}
  \int\dd n\,\Str(\Lambda Q_n).
  \label{eq:continuumaction}
\end{align}
The first term penalizes spatial variations of \(Q\); the second gives a
small mass to retarded--advanced fluctuations.

\subsection{Gaussian propagator and observable insertions}
\label{subsec:propagator}

The calculation is organized in three parallel steps.  We first derive the
Gaussian propagator shared by all observables, then evaluate the local IPR
insertion, and finally evaluate the global spectral insertion.

\subsubsection{From the nonlinear \(Q\) field to the Gaussian propagator}

We expand around the uniform saddle \(Q=\Lambda\) by introducing a tangent
field \(W\) through the rational parametrization commonly used in the
supersymmetric sigma model~\cite{EfetovBook,MirlinReview2000,
AltlandEtAl2015}
\begin{equation}
  Q
  =
  \Lambda
  \left(
    1+\frac{W}{2}
  \right)
  \left(
    1-\frac{W}{2}
  \right)^{-1}.
  \label{eq:QW_rational}
\end{equation}
The condition
\begin{equation}
  \{W,\Lambda\}=0
  \label{eq:Wanticommute}
\end{equation}
ensures \(Q^2=1\).  Expanding Eq.~\eqref{eq:QW_rational} gives
\begin{equation}
  Q
  =
  \Lambda
  \left(
    1+W+\frac12W^2+O(W^3)
  \right).
  \label{eq:QWexpansion}
\end{equation}
Because \(W\) anticommutes with \(\Lambda\), it has no retarded--retarded or
advanced--advanced block.  We therefore write
\begin{equation}
  W(q)
  \equiv
  \begin{pmatrix}
    0&B(q)\\
    \widetilde B(q)&0
  \end{pmatrix}_{\AR}.
  \label{eq:W_B_blocks}
\end{equation}
Thus \(B\) and \(\widetilde B\) are not additional fields: they are the two
off-diagonal blocks of \(W\).

Inserting Eq.~\eqref{eq:QWexpansion} into
Eq.~\eqref{eq:continuumaction}, keeping exactly two powers of \(W\), and
using Eq.~\eqref{eq:W_B_blocks}, we obtain
\begin{align}
  \cS_0
  &=
  \frac1{t_0}
  \int\frac{\dd q}{2\pi}
  \str[B(q)\widetilde B(-q)]
  \nonumber\\
  &\quad\times
  \left[
    |q|^\sigma+t_0(\epsilon-\iu\omega)
  \right].
  \label{eq:S0B}
\end{align}
We denote this quadratic action by \(\cS_0\) because it defines the Gaussian
measure used for every contraction below.

The inverse of the quadratic kernel is the propagator.  We define
\begin{align}
  \Pi(q,\omega)
  &\equiv
  \av{
    B(q)\widetilde B(-q)
  }_0
  \nonumber\\
  &=
  \frac{t_0}{
    |q|^\sigma+t_0(\epsilon-\iu\omega)
  }
  \nonumber\\
  &=
  \frac1{
    D_\sigma|q|^\sigma+\epsilon-\iu\omega
  }.
  \label{eq:Piqw}
\end{align}
Here \(\av{\cdots}_0\) means averaging with the Gaussian weight
\(\exp(-\cS_0)\).  For the static IPR calculation we set
\(\omega=0\), remove \(\epsilon\) after the zero mode has been treated
exactly, and obtain
\begin{equation}
  \Pi(q,0)=\frac{t_0}{|q|^\sigma},
  \qquad q\neq0.
  \label{eq:Piq0}
\end{equation}
The restriction \(q\neq0\) is essential: the uniform mode belongs to the CUE integral and is not treated by Gaussian expansion.  The zero-mode separation and the class-A Wick identities used in the following source expansion are collected in Appendix~\ref{app:sources}.

We have therefore completed the logical transition:
\(Q\) is the nonlinear field, \(W\) is its small tangent fluctuation, and
\(B,\widetilde B\) are the two off-diagonal blocks whose Gaussian inverse is
\(\Pi\).  The following observable calculations use this single propagator.

\subsubsection{IPR moments and fluctuations}

We first evaluate the uniform mode, denoted by the subscript \(0\).
The separation of the exact zero mode from Gaussian nonzero modes follows
the zero-dimensional reduction of the sigma model
\cite{Efetov1983,MirlinReview2000,AltlandEtAl2015}.
Because this mode contains no spatial dependence, its integral is exactly
the class-A random-vector integral.  It gives the zero-mode
moment
\begin{equation}
  \av{I_p}_0=\frac{p!}{N^{p-1}}.
  \label{eq:Ip0}
\end{equation}
This equation fixes the class-A random-vector normalization before any
nonzero momentum mode is contracted~\cite{FyodorovMirlin1995,
MirlinReview2000,EversMirlin2008}.
We now expand the local source insertion in the nonzero field \(W\).
The first correction contains two \(W\) fields.  Choosing which two of the
\(p\) intensity factors supply these fields gives
\(\binom p2=p(p-1)/2\) unordered choices.  Translation invariance makes the
contraction local:
\begin{equation}
  \Pi_{nn}
  \equiv
  \frac1N
  \sum_{q\neq0}\Pi(q,0).
  \label{eq:Pinn_definition}
\end{equation}
Multiplying the zero-mode result by this contraction gives
\begin{align}
  \av{I_p}
  &=\frac{p!}{N^{p-1}}
  \left[1+\frac{p(p-1)}{2N}\sum_{q\neq0}\Pi(q,0)+O(\Pi^2)\right].
  \label{eq:Ip_one_loop}
\end{align}
At the next order there are two logically different contributions:
two independent first-order contractions, proportional to
\(\Pi_{nn}^2\), and a connected two-propagator contraction.  We introduce
\begin{equation}
  \Pi_2
  \equiv
  \frac1{N^2}
  \sum_{q\neq0}\Pi(q,0)^2
  \label{eq:Pi2_definition}
\end{equation}
for the connected momentum sum.  The organization into disconnected exponentiating pieces and connected
composite-operator corrections follows the sigma-model treatment of IPR
moments~\cite{HofWegner1986,Wegner1987I,Wegner1987II,
RushkinOssipovFyodorov2011}.  The moment through second order is
\begin{align}
  \av{I_p}
  &=\frac{p!}{N^{p-1}}
  \Bigg[1+\frac{p(p-1)}2\Pi_{nn}
  +\frac{p^2(p-1)^2}{8}\Pi_{nn}^2
  \nonumber\\
  &\hspace{15mm}-\frac{p(p-1)(2p-1)}4\Pi_2+O(t_0^3)\Bigg],
  \label{eq:Ip_second_order_main}
\end{align}
The term proportional to \(\Pi_{nn}^2\) exponentiates the first-order
correction, whereas the \(\Pi_2\) term changes the connected scaling of the
local moment.
To compute the sample-to-sample fluctuation of the ordinary IPR, we insert
the \(p=2\) source operator twice and keep only contractions connecting the
two insertions.  Two diffuson lines are required, and the class-A internal
index contraction gives
\begin{equation}
  \frac{\operatorname{var}I_2}{\av{I_2}^2}
  =\frac8{N^2}\sum_{q\neq0}\Pi(q,0)^2+O(\Pi^3).
  \label{eq:IPRvariance}
\end{equation}
The source combinatorics leading to
Eqs.~\eqref{eq:Ip_second_order_main} and \eqref{eq:IPRvariance} is given in
Appendix~\ref{app:sources}, specifically Sec.~\ref{app:contractions}.

\subsubsection{Two-level spectral statistics}

We now repeat the same expansion for the global two-level source.  The
frequency \(\omega\) appearing in the sigma-model action is the shift of one
retarded or advanced sector.  Since the two sectors are shifted in opposite
directions, the physical separation is \(\Omega=2\omega\).  Therefore
\begin{equation}
  \omega=\frac{\Omega}{2}=\frac{s\Delta}{2}.
  \label{eq:omega_half_separation}
\end{equation}
This factor of two is needed when the propagator is converted into an
unfolded spectral formula.

On a ring of \(N\) momentum sites, we introduce the discrete wave numbers
\begin{equation}
  q_m\equiv\frac{2\pi m}{N},
  \qquad m\in\mathbb Z,
  \label{eq:qm_definition}
\end{equation}
and the relaxation eigenvalues
\begin{equation}
  \varepsilon_m
  \equiv
  D_\sigma|q_m|^\sigma.
  \label{eq:epsilon_m_definition}
\end{equation}
The mode \(m=0\) is the uniform mode and must be integrated exactly.  The
modes \(m\neq0\) are the Gaussian modes described by \(\Pi\).

\paragraph{Uniform mode: CUE.}

The exact uniform integral gives the CUE cluster function of
Dyson's circular unitary ensemble~\cite{Dyson1962,MehtaBook,HaakeBook,
ZirnbauerCircular1996,ZirnbauerPair1999}
\begin{equation}
  Y_{\rm CUE}^{(N)}(s)
  \equiv
  \left[
    \frac{\sin(\pi s)}{N\sin(\pi s/N)}
  \right]^2.
  \label{eq:Y_CUE_finite}
\end{equation}
Consequently,
\begin{align}
  R_{2,\rm CUE}^{(N)}(s)
  &=1-Y_{\rm CUE}^{(N)}(s),
  \label{eq:R2_CUE_finite}\\
  \mathcal K_{2,\rm CUE}^{(N)}(s)
  &=\delta(s)-Y_{\rm CUE}^{(N)}(s).
  \label{eq:K2_CUE_finite}
\end{align}
For fixed \(s\) and \(N\to\infty\),
\begin{equation}
  Y_{\rm CUE}^{(N)}(s)
  \longrightarrow
  \left(\frac{\sin\pi s}{\pi s}\right)^2.
  \label{eq:CUE_sine_limit}
\end{equation}
This is the local sine-kernel limit of CUE.

\paragraph{First nonzero-mode contraction.}

Each global density derivative starts at second order in \(W\).  Connecting
one retarded density vertex to one advanced density vertex therefore
produces two diffusons.  We introduce
\begin{equation}
  \mathcal C_N
  \equiv
  \frac1{N^2}
  \sum_{m\neq0}
  \Pi(q_m,0)^2.
  \label{eq:CN_definition}
\end{equation}
The sum in Eq.~\eqref{eq:CN_definition} includes both signs of \(m\).  With
this convention the class-A contraction gives, inside the zero-dimensional
window,
\begin{align}
  R_2(s)
  &=
  R_{2,\rm CUE}^{(N)}(s)
  \nonumber\\
  &\quad+
  \frac{\mathcal C_N}{8}
  \frac{\dd^2}{\dd s^2}
  \left[
    s^2R_{2,\rm CUE}^{(N)}(s)
  \right]
  +O(\mathcal C_N^2).
  \label{eq:R2low}
\end{align}
The factor \(1/8\) is the unitary-class source contraction for a momentum
sum over both \(q_m\) and \(-q_m\); analogous two-diffuson corrections are
obtained in the supersymmetric calculations of
Refs.~\cite{AltshulerShklovskii1986,AndreevAltshuler1995,
MirlinReview2000}.  The global source contraction and its normalization are
given in Appendix~\ref{app:correlator}.

Above the zero-dimensional window, the exact CUE mode is no longer the only
soft contribution.  Expanding directly about \(Q=\Lambda\) gives the
connected nonzero-mode correction
\begin{equation}
  \delta\mathcal K_2^{\rm nz}(s)
  =
  \frac{\Delta^2}{4\pi^2}
  \Re
  \sum_{m\neq0}
  \Pi\!\left(q_m,\frac{s\Delta}{2}\right)^2.
  \label{eq:R2correction}
\end{equation}
The ordinary IPR contains one static diffuson at first order, while the
spectral correlator contains two frequency-dependent diffusons.

\paragraph{Class-A spectral determinant.}

For later use we package all nonzero relaxation modes into the spectral
determinant introduced in the beyond-random-matrix treatment of spectral
correlations~\cite{AndreevAltshuler1995,AgamAltshulerAndreev1995,
MirlinReview2000}
\begin{equation}
  \mathcal D_N(s)
  \equiv
  \frac1{s^2}
  \prod_{m\neq0}
  \left[
    1+
    \frac{s^2\Delta^2}{4\varepsilon_m^2}
  \right]^{-1}.
  \label{eq:spectral_determinant}
\end{equation}
The factor \(1/4\) follows from Eq.~\eqref{eq:omega_half_separation}.  The
Gaussian-mode product and its equivalence to the direct two-diffuson
contraction are derived in Appendix~\ref{app:correlator}.  The class-A
normal saddle gives
\begin{equation}
  \mathcal K_{2,\rm pert}(s)
  =
  -\frac1{4\pi^2}
  \frac{\dd^2}{\dd s^2}
  \ln\mathcal D_N(s),
  \label{eq:K2_pert_D}
\end{equation}
whereas the nonstandard Andreev--Altshuler saddle
\cite{AndreevAltshuler1995} gives
\begin{equation}
  \mathcal K_{2,\rm osc}(s)
  =
  \frac{\cos(2\pi s)}{2\pi^2}
  \mathcal D_N(s).
  \label{eq:K2_osc_D}
\end{equation}
The relaxation spectrum \(\varepsilon_m\) in these equations is not taken
from a random-matrix model: it is the spectrum of the fractional rotor
kernel derived in Sec.~\ref{sec:fourier}.  The standard class-A normal and nonstandard saddle formulas are used only
to perform the remaining nonlinear zero-mode integral
\cite{AndreevAltshuler1995,AgamAltshulerAndreev1995,MirlinReview2000,
ZirnbauerPair1999}.

\paragraph{Number variance and spectral compressibility.}

To measure correlations over an interval containing on average \(L\) levels,
we introduce the counting variable
\begin{equation}
  \mathcal N_L(x)
  \equiv
  \int_x^{x+L}\varrho(u)\dd u.
  \label{eq:number_count_definition}
\end{equation}
Its variance is the standard level-number variance used in metallic and
critical spectral statistics~\cite{AltshulerShklovskii1986,
Kravtsov1996,ChalkerKravtsovLerner1996,MirlinReview2000}:
\begin{align}
  \Sigma_N^2(L)
  &\equiv
  \av{[\mathcal N_L-\av{\mathcal N_L}]^2}
  \nonumber\\
  &=
  \int_{-L}^{L}
  (L-|s|)\mathcal K_{2,N}(s)\dd s
  \nonumber\\
  &=
  L-2\int_0^L(L-s)Y_{2,N}(s)\dd s.
  \label{eq:number_variance_from_Y}
\end{align}
We define the thermodynamic spectral compressibility by taking the system
size to infinity before the spectral window:
\begin{equation}
  \chi
  \equiv
  \lim_{L\to\infty}
  \lim_{N\to\infty}
  \frac{\Sigma_N^2(L)}{L}.
  \label{eq:chi_definition}
\end{equation}
If the thermodynamic cluster function is integrable, Eq.~\eqref{eq:number_variance_from_Y}
gives
\begin{equation}
  \chi
  =
  1-
  \int_{-\infty}^{\infty}Y_2(s)\dd s.
  \label{eq:chi_cluster_integral}
\end{equation}
The order of limits in Eq.~\eqref{eq:chi_definition} is essential in the
localized phase because finite spectra obey a global level-number sum rule.

\subsection{One-loop cancellation and two-loop stiffness}
\label{subsec:RG}

The action renormalization derived here is common to both observable
sectors.  In each phase below we therefore use the same order:
RG flow, IPR moments, and spectral statistics.

The flow is organized by integrating a thin momentum shell and restoring the
cutoff, in the spirit of the scaling theory of localization and its
nonlinear-sigma-model implementation~\cite{AbrahamsEtAl1979,Wegner1979,
Hikami1981,EfetovBook}.  The tree term comes from rescaling the nonlocal
kinetic operator; loop terms come from connected fast-mode cumulants.

The one-loop cancellation concerns the stiffness self-energy, not the direct
source corrections above.  The background-field organization and the
loop-order structure of the unitary nonlinear sigma model follow
Refs.~\cite{Hikami1981,HofWegner1986,Wegner1987I,Wegner1987II,
MirlinReview2000}. Expanding the kinetic action gives
\begin{align}
  \cS_4
  &=\frac1{32t}\sum_{n,m}\mathcal U_{nm}\Str[
    W_n^3W_m+W_nW_m^3-W_n^2W_m^2].
  \label{eq:S4_main}
\end{align}
After splitting \(W=W_<+W_>\), where \(W_<\) contains the slow external
modes below the running shell and \(W_>\) contains the fast modes inside the
thin momentum shell being integrated out, the cubic--linear terms give
$+2t\mathcal I_\sigma(q)$ and the two cross-contractions of
$-W_n^2W_m^2$ give $-2t\mathcal I_\sigma(q)$, with
\begin{equation}
  \mathcal I_\sigma(q)=\frac12\int_{>}\frac{\dd k}{2\pi}
  \frac{|q+k|^\sigma-|k|^\sigma}{|k|^\sigma}.
  \label{eq:loopfunction}
\end{equation}
Hence
\begin{equation}
  \Sigma^{(1)}(q)=0.
  \label{eq:oneloopcancel}
\end{equation}
This is the standard class-A cancellation: broken time reversal removes the
cooperon, so the weak-localization correction to the stiffness is absent at
one loop~\cite{Hikami1981,EfetovBook,EversMirlin2008}. It does not set Eqs.~\eqref{eq:Ip_one_loop} or
\eqref{eq:R2correction} to zero.

At two loops,
\begin{equation}
  \delta\cS_2^{(2)}=\av{\cS_6}_{>}
  -\frac12\av{\cS_4^2}_{>,c}+\delta\cS_J^{(2)}.
  \label{eq:two_loop_cumulant}
\end{equation}
The full vertices and contracted integrals are displayed in
Appendix~\ref{app:two_loop}.  In class A the one-loop stiffness term is
absent and the first interference contribution to the beta function occurs
at two loops~\cite{Hikami1981,Wegner1987II,MirlinReview2000,
EversMirlin2008}.  Appendix~\ref{app:two_loop} explicitly shows how the
two-loop quadratic action first renormalizes the inverse stiffness, how the
cutoff-restoring rescaling supplies the engineering term, and how the two
pieces combine.  After matching the conventional kinetic normalization to
the fractional rotor kernel, one obtains
\begin{equation}
  \frac{\dd t}{\dd\ell}=(\sigma-1)t+\frac{t^3}{2\pi^2}+O(t^5),
  \qquad \sigma>1.
  \label{eq:beta_localized_A}
\end{equation}

\subsection{Localized phase: $\sigma>1$}
\label{subsec:localized_phase}

\subsubsection{RG flow and localization length}

Let $\delta=\sigma-1$ and $a_A=1/(2\pi^2)$. Setting $y=t^{-2}$ gives
\begin{equation}
  \frac{\dd y}{\dd\ell}=-2\delta y-2a_A.
\end{equation}
Thus
\begin{equation}
  \frac1{t^2(\ell)}=\e^{-2\delta\ell}
  \left[\frac1{t_0^2}+\frac{a_A}{\delta}\right]-\frac{a_A}{\delta},
  \label{eq:tflowlocalized}
\end{equation}
and
\begin{equation}
  \frac{\xi_A}{L_0}=\left[1+\frac{2\pi^2(\sigma-1)}{t_0^2}\right]^{1/[2(\sigma-1)]}.
  \label{eq:xiA}
\end{equation}
The integration of the differential equation and the definition of the
strong-coupling stopping scale are also given explicitly in
Appendix~\ref{app:two_loop}.

\subsubsection{IPR moments}

The renormalized IPR is obtained by multiplicatively renormalizing the
local moment operator, as in the composite-operator treatments of
Refs.~\cite{HofWegner1986,Wegner1987I,Wegner1987II,
RushkinOssipovFyodorov2011}.  We introduce the source factor $Z_p$ by
\begin{align}
  \av{I_p(L)}&=\frac{p!}{L^{p-1}}Z_p(L),
  \label{eq:IpRGdef}\\
  \frac{\dd\ln Z_p}{\dd\ell}&=\gamma_p[t(\ell)],
  \qquad \gamma_p(t)=\frac{p(p-1)}{2\pi}t+O(t^3).
  \label{eq:gammaP}
\end{align}
Here $Z_p$ is the multiplicative renormalization factor of the composite
local source operator; it is not the partition function \(\cZ\).  The
explicit slow--fast split of the source vertex, the shell contraction that
produces \(\gamma_p\), and the conversion from the shell recursion to
Eq.~\eqref{eq:gammaP} are derived in Appendix~\ref{app:two_loop}.
Using the beta function,
\begin{align}
  \ln\frac{Z_p(L)}{Z_p(L_0)}
  &=\int_{t_0}^{t(L)}\frac{\gamma_p(t)}{\delta t+a_At^3}\dd t
  \nonumber\\
  &=\frac{p(p-1)}{2\pi\sqrt{a_A\delta}}
  \Bigg[
  \arctan\!\left(t(L)\sqrt{\frac{a_A}{\delta}}\right)
  \nonumber\\
  &\hspace{18mm}-
  \arctan\!\left(t_0\sqrt{\frac{a_A}{\delta}}\right)
  \Bigg].
  \label{eq:ZpLocalized}
\end{align}
This is controlled for $L\ll\xi_A$. At $L\gtrsim\xi_A$,
$\av{I_p}\sim\xi_A^{1-p}$.

\subsubsection{Spectral statistics: CUE-to-Poisson crossover}

The RG flow first determines the localization length \(\xi_A\).  For
\(N\ll\xi_A\), the whole rotor is one coherent localization volume and the
uniform integral gives finite-\(N\) CUE.  For \(N\gg\xi_A\), localized eigenstates occupy asymptotically independent
localization volumes, as in the standard sigma-model description of the
localized regime~\cite{EfetovBook,MirlinReview2000,EversMirlin2008}.  Write
\begin{equation}
  M\equiv\frac{N}{\xi_A}
  \label{eq:number_localization_blocks}
\end{equation}
for the number of localization volumes; taking \(M\) integer is only a
notational simplification.  Each volume contributes a statistically
independent spectrum with density \(1/M\) in the globally unfolded units.
The CUE cluster function of one volume, unfolded with its own mean spacing,
is
\begin{equation}
  Y_{\rm CUE}^{(\xi_A)}(u)
  =
  \left[
    \frac{\sin(\pi u)}{\xi_A\sin(\pi u/\xi_A)}
  \right]^2.
  \label{eq:local_block_CUE_cluster}
\end{equation}
Only two levels belonging to the same volume are correlated.  Summing the
\(M\) independent connected contributions gives
\begin{align}
  Y_{2,N}^{\rm loc}(s)
  &=
  M\left(\frac1M\right)^2
  Y_{\rm CUE}^{(\xi_A)}\!\left(\frac{s}{M}\right)
  \nonumber\\
  &=
  \frac1M
  Y_{\rm CUE}^{(\xi_A)}\!\left(\frac{s}{M}\right)
  \nonumber\\
  &=
  \frac{\xi_A}{N}
  \left[
    \frac{
      \sin(\pi s\xi_A/N)
    }{
      \xi_A\sin(\pi s/N)
    }
  \right]^2.
  \label{eq:localized_cluster_superposition}
\end{align}
Therefore the off-diagonal pair correlator is
\begin{equation}
  R_{2,N}^{\rm loc}(s)
  =
  1-Y_{2,N}^{\rm loc}(s).
  \label{eq:R2_localized_finiteN}
\end{equation}
For every fixed unfolded separation \(s\),
\begin{align}
  \lim_{N/\xi_A\to\infty}
  Y_{2,N}^{\rm loc}(s)
  &=
  \lim_{M\to\infty}
  \frac1M
  Y_{\rm CUE}^{(\xi_A)}(s/M)
  =0,
  \nonumber\\
  \lim_{N/\xi_A\to\infty}
  R_{2,N}^{\rm loc}(s)&=1.
  \label{eq:R2_to_Poisson}
\end{align}
Thus the connected part \(R_2-1\) vanishes and the thermodynamic local
statistics is Poisson.

The same conclusion follows directly from the number variance.  The
linear Poisson variance and the vanishing of connected correlations between
independent localization volumes are the conventional localized limits of
the sigma model~\cite{EfetovBook,MirlinReview2000,EversMirlin2008}.  Since
\(0\leq Y_{\rm CUE}^{(\xi_A)}(u)\leq1\),
\begin{align}
  0
  &\leq
  2\int_0^L(L-s)Y_{2,N}^{\rm loc}(s)\dd s
  \nonumber\\
  &\leq
  \frac{L^2}{M}.
  \label{eq:localized_number_variance_bound}
\end{align}
Taking \(N\to\infty\) first at fixed \(L\) makes the correction vanish:
\begin{align}
  \lim_{N\to\infty}\Sigma_N^2(L)
  &=L,
  \nonumber\\
  \chi_{\rm loc}&=1.
  \label{eq:chi_localized}
\end{align}
The running propagator description is valid up to
\(L_*=\min(N,L_\omega)\), with
\(L_\omega=(D_\sigma/|\omega|)^{1/\sigma}\).  Once
\(L_*\gtrsim\xi_A\), the block decomposition above supplies the
nonperturbative continuation of the RG and makes the CUE-to-Poisson
crossover explicit.

\subsection{Critical phase: $\sigma=1$}
\label{subsec:critical_phase}

\subsubsection{Marginal RG flow}

At the exact \(|q|\) kernel, the tree-level scaling is marginal.  This is
the weak-coupling critical nonlocal sigma-model regime studied for PRBM
ensembles in Refs.~\cite{MirlinEtAl1996,MirlinEvers2000,
RushkinOssipovFyodorov2011}.  In the present class-A rotor,
\begin{equation}
  \frac{\dd t}{\dd\ell}=0.
  \label{eq:critical_t_flow}
\end{equation}
Thus the critical coupling remains equal to its microscopic value \(t_0\).
The distinction between the localized-side shell renormalization and the
exactly marginal nonanalytic \(|q|\) coefficient is explained at the end of
Appendix~\ref{app:two_loop}.

\subsubsection{IPR moments and multifractal dimensions}

The local moment operator continues to renormalize even though the
stiffness is marginal:
\begin{equation}
  \frac{\dd\ln Z_p}{\dd\ell}
  =\frac{p(p-1)}{2\pi}t_0+O(t_0^3).
\end{equation}
Therefore
\begin{align}
  \av{I_p}
  &=\frac{p!}{N^{p-1}}
  \left(\frac{N}{L_0}\right)^{p(p-1)t_0/(2\pi)+O(t_0^3)},
  \nonumber\\
  d_p&=1-\frac{pt_0}{2\pi}+O(t_0^3).
  \label{eq:dpA}
\end{align}
The independent $O(t_0^2)$ anomalous dimension vanishes in class A,
consistent with the unitary weak-multifractality calculation of
Ref.~\cite{RushkinOssipovFyodorov2011}; the first nonparabolic term is
$O(t_0^3)$ and depends on the full rotor kernel.  Its explicit
kernel-dependent coefficient is recorded in Appendix~\ref{app:third_order}.

\subsubsection{Spectral statistics and explicit compressibility}

At \(\sigma=1\), the nonzero relaxation eigenvalues on the ring are
\begin{equation}
  \varepsilon_m
  =D_1\frac{2\pi|m|}{N}
  =\Delta\frac{|m|}{t_0},
  \qquad m\neq0,
  \label{eq:critical_relaxation_eigenvalues}
\end{equation}
because \(D_1=t_0^{-1}\).  The combination of the normal and
Andreev--Altshuler saddles follows the nonperturbative spectral-correlation
construction of Refs.~\cite{AndreevAltshuler1995,
AgamAltshulerAndreev1995,MirlinEtAl1996,MirlinReview2000}.  Combining the
relaxation eigenvalue with \(\omega=s\Delta/2\), every determinant factor
becomes
\begin{equation}
  1+
  \frac{s^2\Delta^2}{4\varepsilon_m^2}
  =
  1+
  \left(\frac{s t_0}{2m}\right)^2.
  \label{eq:critical_determinant_factor}
\end{equation}
The modes \(m\) and \(-m\) are degenerate, so
\begin{align}
  \mathcal D_{\rm crit}(s)
  &=
  \frac1{s^2}
  \prod_{m=1}^{\infty}
  \left[
    1+\left(\frac{s t_0}{2m}\right)^2
  \right]^{-2}.
  \label{eq:Dcrit_product}
\end{align}
We now use the Euler product
\begin{equation}
  \frac{\sinh(\pi z)}{\pi z}
  =
  \prod_{m=1}^{\infty}
  \left(1+\frac{z^2}{m^2}\right)
  \label{eq:sinh_Euler_product}
\end{equation}
with \(z=s t_0/2\).  This gives
\begin{equation}
  \mathcal D_{\rm crit}(s)
  =
  \frac1{s^2}
  \left[
    \frac{x}{\sinh x}
  \right]^2,
  \qquad
  x\equiv\frac{\pi t_0s}{2}.
  \label{eq:Dcrit_closed}
\end{equation}
Substituting Eq.~\eqref{eq:Dcrit_closed} into the normal- and
Andreev--Altshuler-saddle contributions,
Eqs.~\eqref{eq:K2_pert_D} and \eqref{eq:K2_osc_D}, and simplifying the two
terms gives
\begin{align}
  \mathcal K_{2,\rm crit}(s)
  &=
  \delta(s)-Y_{2,\rm crit}(s),
  \label{eq:K2_critical_closed}\\
  Y_{2,\rm crit}(s)
  &=
  \left(\frac{\sin\pi s}{\pi s}\right)^2
  \left(\frac{x}{\sinh x}\right)^2,
  \label{eq:Y2_critical_closed}\\
  R_{2,\rm crit}(s)
  &=
  1-
  \left(\frac{\sin\pi s}{\pi s}\right)^2
  \left(\frac{x}{\sinh x}\right)^2.
  \label{eq:R2_critical_closed}
\end{align}
The two limiting statistics are now obtained directly from the formula.  At
weak critical coupling,
\begin{equation}
  \lim_{t_0\to0}\frac{x}{\sinh x}=1,
\end{equation}
so
\begin{equation}
  \lim_{t_0\to0}R_{2,\rm crit}(s)
  =
  1-
  \left(\frac{\sin\pi s}{\pi s}\right)^2,
  \label{eq:critical_to_CUE}
\end{equation}
which is CUE.  In the formal strong-coupling limit, for fixed \(s\neq0\),
\begin{equation}
  \frac{x}{\sinh x}\sim2x\e^{-x}\longrightarrow0,
\end{equation}
and therefore
\begin{equation}
  \lim_{t_0\to\infty}R_{2,\rm crit}(s)=1,
  \label{eq:critical_to_Poisson}
\end{equation}
which is Poisson.  The controlled weak-coupling rotor theory uses the first
part of this interpolation; the second limit displays how the pure scaling
sigma model loses spectral rigidity.

We next compute the compressibility instead of quoting it.  Linear
critical number variance and its relation to multifractal wave-function
correlations were developed in Refs.~\cite{KravtsovEtAl1994,
Kravtsov1996,ChalkerKravtsovLerner1996}; the
limitations of extending the weak-multifractality relation to arbitrary
coupling are discussed in Refs.~\cite{EversMirlin2000PRL,
MirlinEvers2000}.  Introduce
\begin{equation}
  a\equiv\frac{\pi t_0}{2}.
  \label{eq:a_critical_definition}
\end{equation}
Equation~\eqref{eq:Y2_critical_closed} becomes
\begin{equation}
  Y_{2,\rm crit}(s)
  =
  \frac{a^2}{\pi^2}
  \frac{\sin^2(\pi s)}{\sinh^2(as)}.
  \label{eq:Ycritical_for_integral}
\end{equation}
Using Eq.~\eqref{eq:chi_cluster_integral},
\begin{equation}
  \chi_{\rm crit}
  =
  1-
  \frac{a^2}{\pi^2}
  \int_{-\infty}^{\infty}
  \frac{\sin^2(\pi s)}{\sinh^2(as)}\dd s.
  \label{eq:chi_critical_integral}
\end{equation}
For \(s>0\), expand
\begin{equation}
  \frac1{\sinh^2(as)}
  =
  4\sum_{n=1}^{\infty}n\e^{-2nas}.
  \label{eq:csch2_series}
\end{equation}
The elementary Laplace integral is
\begin{equation}
  \int_0^\infty
  \e^{-cs}\sin^2(\pi s)\dd s
  =
  \frac{2\pi^2}{c(c^2+4\pi^2)}.
  \label{eq:laplace_sin2}
\end{equation}
Setting \(c=2na\), summing over \(n\), and using
\begin{equation}
  \sum_{n=1}^{\infty}
  \frac1{n^2a^2+\pi^2}
  =
  \frac1{2\pi^2}
  \left[
    \frac{\pi^2}{a}
    \coth\left(\frac{\pi^2}{a}\right)
    -1
  \right],
  \label{eq:coth_sum_identity}
\end{equation}
we obtain
\begin{equation}
  \int_{-\infty}^{\infty}
  \frac{\sin^2(\pi s)}{\sinh^2(as)}\dd s
  =
  \frac{\pi^2}{a^2}
  \coth\left(\frac{\pi^2}{a}\right)
  -\frac1a.
  \label{eq:critical_integral_evaluated}
\end{equation}
Substitution into Eq.~\eqref{eq:chi_critical_integral} gives the explicit
pure-scaling result
\begin{equation}
  \chi_{\rm crit}^{\rm scaling}(t_0)
  =
  1-
  \coth\left(\frac{2\pi}{t_0}\right)
  +\frac{t_0}{2\pi}.
  \label{eq:chi_critical_closed}
\end{equation}
For \(t_0\ll1\),
\begin{equation}
  \chi_{\rm crit}^{\rm scaling}
  =
  \frac{t_0}{2\pi}
  -2\e^{-4\pi/t_0}
  +O(\e^{-8\pi/t_0}).
  \label{eq:chi_critical_weak}
\end{equation}
The full deterministic rotor kernel and interaction vertices can modify the
nonuniversal terms beyond the pure \(|q|\) scaling approximation.  The
controlled perturbative statement is therefore
\begin{equation}
  \chi_{\rm QKR}
  =
  \frac{t_0}{2\pi}+O(t_0^3).
  \label{eq:chi_QKR_perturbative}
\end{equation}
Using \(d_2=1-t_0/\pi+O(t_0^3)\), this becomes, to the same order,
\begin{equation}
  \chi_{\rm QKR}
  =
  \frac{1-d_2}{2}+O(t_0^3).
  \label{eq:chi_d2_relation}
\end{equation}
The relation is a leading weak-multifractality result, not an exact identity
at arbitrary coupling.

\subsection{Extended phase: $0<\sigma<1$}
\label{subsec:extended_phase}

\subsubsection{RG flow}

The Ward-identity subtraction and nonlocal power counting follow the
long-range sigma-model RG of Refs.~\cite{MirlinEtAl1996,
MirlinReview2000}.  Appendix~\ref{app:two_loop} defines the one-particle
irreducible self-energy and derives the loop counting explicitly.  In one
dimension its infrared $L$-loop contribution obeys
\begin{equation}
  \Sigma_L(q,0)
  \sim
  t^{L-1}|q|^{\sigma+L(1-\sigma)},
  \qquad 0<\sigma<1,
  \label{eq:selfenergy_scaling_extended}
\end{equation}
so, relative to the bare inverse propagator
$\Pi_0^{-1}(q,0)=|q|^\sigma/t$,
\begin{equation}
  \frac{\Sigma_L(q,0)}{\Pi_0^{-1}(q,0)}
  \sim
  t^L|q|^{L(1-\sigma)}
  \longrightarrow0
  \qquad (q\to0).
  \label{eq:selfenergy_relative_extended}
\end{equation}
Thus loop self-energies cannot renormalize the nonanalytic
$|q|^\sigma$ stiffness in the extended phase.  In class A the one-loop
coefficient vanishes identically, so the first possible nonzero term is the
two-loop correction
$\Sigma_2(q,0)\sim t|q|^{2-\sigma}$, which is still more irrelevant than
the bare kernel.  Thus
\begin{equation}
  \frac{\dd t}{\dd\ell}=(\sigma-1)t,
  \qquad t(L)=t_0(L/L_0)^{\sigma-1}.
  \label{eq:betaextended}
\end{equation}
\subsubsection{IPR moments and fluctuations}

The source integration is
\begin{align}
  \ln\frac{Z_p(L)}{Z_p(L_0)}
  &=\frac{p(p-1)}{2\pi}\int_0^{\ln(L/L_0)}
  t_0\e^{-(1-\sigma)\ell}\dd\ell
  \nonumber\\
  &=\frac{p(p-1)t_0}{2\pi(1-\sigma)}
  \left[1-(L/L_0)^{\sigma-1}\right].
  \label{eq:Zpextended}
\end{align}
It approaches a finite constant, so $\av{I_p}\propto N^{1-p}$.
The IPR variance scales as
\begin{equation}
  \frac{\operatorname{var}I_2}{\av{I_2}^2}\propto
  \begin{cases}
    N^{-2+2\sigma},&1/2<\sigma<1,\\
    N^{-1}\ln N,&\sigma=1/2,\\
    N^{-1},&0<\sigma<1/2.
  \end{cases}
\end{equation}
The boundary \(\sigma=1/2\) is only a two-diffuson convergence threshold.

\subsubsection{Spectral statistics: convergence to CUE}

For a finite ring, the first nonzero relaxation rate is the fractional
Thouless energy
\begin{equation}
  E_{\rm Th}(N)
  \equiv
  D_\sigma\left(\frac{2\pi}{N}\right)^\sigma.
  \label{eq:Eth_fractional}
\end{equation}
We introduce the dimensionless Thouless conductance
\begin{align}
  g_N
  &\equiv
  \frac{E_{\rm Th}(N)}{\Delta}
  \nonumber\\
  &=
  (2\pi)^{\sigma-1}
  D_\sigma N^{1-\sigma}
  \nonumber\\
  &=
  \frac{(2\pi)^{\sigma-1}}{t_0}
  N^{1-\sigma}.
  \label{eq:gN_fractional}
\end{align}
For every \(0<\sigma<1\), \(g_N\to\infty\).  This is the fractional
analogue of the Thouless-number criterion for the crossover from spatial
sigma-model modes to Wigner--Dyson statistics
\cite{AltshulerShklovskii1986,KravtsovMirlin1994,
AndreevAltshuler1995,MirlinReview2000}.  At fixed unfolded separation
\(s\),
\begin{equation}
  \frac{\Omega}{E_{\rm Th}}
  =
  \frac{s}{g_N}
  \longrightarrow0,
  \label{eq:fixed_s_zero_dimensional}
\end{equation}
so the system enters the zero-dimensional CUE window.

The convergence can be checked directly from the two-diffuson sum.  Using
\(q_m=2\pi m/N\) and \(\Pi(q_m,0)=t_0/|q_m|^\sigma\),
\begin{align}
  \mathcal C_N
  &=
  \frac{2t_0^2}{(2\pi)^{2\sigma}}
  N^{2\sigma-2}
  \sum_{m=1}^{N/2}m^{-2\sigma}.
  \label{eq:CN_explicit_sum}
\end{align}
The large-\(N\) asymptotics is
\begin{equation}
  \mathcal C_N
  \simeq
  \begin{cases}
    \displaystyle
    \frac{2t_0^2\zeta(2\sigma)}{(2\pi)^{2\sigma}}
    N^{-2(1-\sigma)},
    &\frac12<\sigma<1,\\[3mm]
    \displaystyle
    \frac{t_0^2}{\pi}
    \frac{\ln N}{N},
    &\sigma=\frac12,\\[3mm]
    \displaystyle
    \frac{t_0^2}{\pi^{2\sigma}(1-2\sigma)}
    \frac1N,
    &0<\sigma<\frac12.
  \end{cases}
  \label{eq:CN_asymptotic_cases}
\end{equation}
Every branch tends to zero.  Equation~\eqref{eq:R2low} therefore gives
\begin{equation}
  \lim_{N\to\infty}R_2(s)
  =
  1-
  \left(\frac{\sin\pi s}{\pi s}\right)^2,
  \qquad 0<\sigma<1,
  \label{eq:R2_extended_to_CUE}
\end{equation}
for every fixed \(s\).  This is an explicit CUE limit rather than a statement
based only on the sign of the beta function.

The compressibility follows from the CUE cluster integral.  Using
\begin{equation}
  \int_0^\infty
  \left(\frac{\sin\pi s}{\pi s}\right)^2\dd s
  =\frac12,
  \label{eq:sinc_integral}
\end{equation}
we obtain
\begin{equation}
  \chi_{\rm ext}=1-2\times\frac12=0.
  \label{eq:chi_extended}
\end{equation}

At frequencies above the Thouless scale, \(s\gtrsim2g_N\), the
nonzero-mode expression must be used.  Substituting
\(\omega=s\Delta/2\) into Eq.~\eqref{eq:R2correction} and replacing the
sum by an integral gives
\begin{align}
  \delta\mathcal K_2^{\rm nz}(s)
  &\simeq
  \frac{N\Delta^2}{4\pi^3}
  \Re\int_0^\infty
  \frac{\dd q}{[D_\sigma q^\sigma-\iu\omega]^2}
  \nonumber\\
  &=
  \frac{N\Delta^2}{4\pi^3}
  B_\sigma
  D_\sigma^{-1/\sigma}
  |\omega|^{1/\sigma-2},
  \label{eq:AS_tail_explicit}
\end{align}
where
\begin{equation}
  B_\sigma
  =
  -\frac1\sigma
  \Gamma\left(\frac1\sigma\right)
  \Gamma\left(2-\frac1\sigma\right)
  \cos\left(\frac{\pi}{2\sigma}\right).
  \label{eq:Bsigma_again}
\end{equation}
This fractional Altshuler--Shklovskii tail generalizes the diffusive
high-frequency result of Ref.~\cite{AltshulerShklovskii1986} to the
fractional rotor kernel and does not contradict the local CUE
limit: as \(N\to\infty\), \(g_N\to\infty\), so the crossover is pushed to
arbitrarily large unfolded separation.

\section{Universality, comparison with PRBM, and limits}
\label{sec:universality}

The purpose of this section is to identify which parts of the singular-rotor
theory are universal and which retain information about the underlying
Floquet dynamics.  The comparison with power-law random banded matrices
(PRBM) is made only after the rotor action, its coupling, and its observable
insertions have been obtained microscopically.  
\subsection{Criteria for infrared equivalence}

The infrared action derived for the rotor is
\begin{align}
  \cS_{\rm IR}[Q]
  &=
  -\frac{1}{8t_0}
  \int\frac{\dd q}{2\pi}
  |q|^\sigma
  \Str(Q_qQ_{-q})
  \nonumber\\
  &\quad+
  \frac{\epsilon-\iu\omega}{4}
  \int\dd n\,
  \Str(\Lambda Q_n),
  \label{eq:IR_universality_action}
\end{align}
with class-A target manifold
\begin{equation}
  \frac{U(1,1|2)}
  {U(1|1)\times U(1|1)}.
\end{equation}
The same generating functional produces the local moment operator
\(\mathcal O_p\) and the global two-level operator \(\mathcal O_C\).

A class-A PRBM and the present rotor describe the same infrared fixed point
provided that four ingredients agree: the target manifold, the fractional
kinetic exponent \(\sigma\), the normalization of the running coupling, and
the source operators used to define the observables.  Under these
conditions, their beta functions, anomalous dimensions, and infrared
scaling functions can be compared directly
\cite{FyodorovMirlin1991,MirlinEtAl1996,MirlinEvers2000,
RushkinOssipovFyodorov2011}.  Matching only the algebraic decay exponent is
not sufficient, because a change of symmetry class or source normalization
changes the loop coefficients and the observable prefactors.

For a PRBM whose hopping standard deviation obeys
\begin{equation}
  [\operatorname{Var}(H_{nm})]^{1/2}
  \propto
  |n-m|^{-\alpha},
\end{equation}
the probability kernel has exponent
\begin{equation}
  \sigma=2\alpha-1.
  \label{eq:PRBM_alpha_sigma}
\end{equation}
The strict endpoint tail of the positive-\(\beta\) rotor therefore gives
\begin{equation}
  \alpha_{\rm eff}=1+\beta,
  \qquad
  \sigma_{\rm IR}=1+2\beta,
  \label{eq:positive_beta_PRBM_match}
\end{equation}
whereas the logarithmic rotor gives
\begin{equation}
  \alpha_{\rm eff}=1,
  \qquad
  \sigma=1.
  \label{eq:log_PRBM_match}
\end{equation}
For negative \(\beta\), the stationary-phase result gives
\begin{equation}
  \sigma_-=\frac{1}{1-\beta}<1,
  \qquad
  \alpha_{\rm eff}
  =
  \frac{1+\sigma_-}{2}.
  \label{eq:negative_beta_PRBM_match}
\end{equation}

Consequently, whenever the asymptotic window derived in
Sec.~\ref{sec:fourier} is resolved, the rotor reproduces the long-range
Anderson-transition structure:
\begin{align}
  \beta<0
  &:\quad \text{extended},
  \nonumber\\
  \beta=0
  &:\quad \text{critical},
  \nonumber\\
  \beta>0
  &:\quad \text{localized}.
  \label{eq:rotor_PRBM_phase_correspondence}
\end{align}
For \(\beta\geq1/2\), the strict endpoint kernel becomes local or marginally
local, and the qualifications discussed in Sec.~\ref{sec:fourier} apply.
Equation~\eqref{eq:rotor_PRBM_phase_correspondence} is not inferred from
PRBM.  It follows independently from the rotor-derived relaxation kernel and
the RG equations obtained above.

\subsection{Phase-dependent comparison}

The correspondence is strongest in the localized and critical regimes.

For \(\sigma>1\), both theories flow away from weak coupling and develop a
finite localization scale.  Their participation moments become controlled
by one localization volume, and their spectral statistics cross from the
finite-volume class-A form to Poisson statistics between independent
localization volumes.  The microscopic relation between the bare parameters
and \(t_0\), as well as the symmetry-dependent perturbative coefficients,
remains model dependent, but the infrared localization mechanism is the
same.

At \(\sigma=1\), both theories contain the marginal \(|q|\) kinetic term and
the same class-A hierarchy of multifractal and spectral operators.  After
matching the coupling convention, the leading multifractal dimensions,
critical two-level correlations, and spectral compressibility agree.
Higher-order coefficients may depend on the ultraviolet completion of the
kernel, but such dependence does not modify the identification of the
critical fixed point.

The extended phase requires a more qualified statement.  For
\(0<\sigma<1\), both theories flow toward
\begin{equation}
  t(\ell)\longrightarrow0,
\end{equation}
and therefore have the same class-A metallic endpoint:
\begin{equation}
  \av{I_p}\propto N^{1-p},
  \qquad
  R_2(s)\longrightarrow R_{2,\rm CUE}(s).
\end{equation}
The approach to this endpoint, however, need not be identical.  PRBM is a
static ensemble with quenched random hopping, whereas the rotor evolves
through a deterministic Floquet operator whose relaxation spectrum is fixed
by the full function \(1-\widehat P(q)\).  Finite-size corrections,
resonance effects, dynamical crossover functions, the fractional
Altshuler--Shklovskii tail, and the crossover between stationary-phase and
endpoint asymptotics can therefore differ.  The extended rotor and PRBM
share the same infrared fixed point, but generally not the same
preasymptotic dynamics.

A final microscopic distinction should be kept in mind.  PRBM is defined as
an ensemble with quenched random matrix elements, whereas the kicked rotor is
a single deterministic Floquet map.  The center-phase integral in the
present generating functional is a spectral average within that same map,
not a disorder average.  Effective randomization by incommensurate kinetic
phases is the standard kicked-rotor mechanism already present for smooth
kicks~\cite{AltlandZirnbauerQKR1996,AltlandEtAl2015}; the role of the
singularity here is to change the long-distance relaxation kernel from a
short-range one to an algebraic one.  We therefore do not regard the absence
of quenched disorder itself as a separate novelty of the singular model.

\subsection{Weak-coupling domain and limitations}

The RG equations used in this work are perturbative, as in the
weak-coupling treatments of long-range sigma models and multifractal
operators~\cite{MirlinEtAl1996,MirlinReview2000,MirlinEvers2000,
RushkinOssipovFyodorov2011}.  In the present normalization,
\begin{equation}
  t=\frac{1}{D_\sigma}
\end{equation}
is the expansion parameter, and the loop calculation is controlled only
while
\begin{equation}
  t(\ell)\ll1
\end{equation}
and the higher-gradient operators generated under coarse graining remain
irrelevant.

In the localized phase, the perturbative flow can be followed only until
\(t(\ell)\) becomes of order unity.  Defining that stopping scale as
\(\xi\) gives the weak-coupling localization length, but does not describe
the subsequent strong-coupling flow.  The asymptotic Poisson limit is a
consequence of localization, whereas the detailed crossover function and
the numerical prefactor of \(\xi\) are quantitatively controlled only when
the bare coupling is sufficiently small.

At criticality, the multifractal dimensions and compressibility are
controlled small-\(t_0\) expansions.  Higher-order terms become important
at strong critical coupling.  In the extended phase, \(t(\ell)\) decreases,
so the perturbative description becomes increasingly accurate in the
infrared, although a strong-coupling microscopic crossover is not captured.
The transition point \(\sigma=1\), fixed by the scaling dimension of the
fractional kinetic operator, is therefore more robust than the particular
weak-coupling coefficients entering \(\xi\), \(d_p\), and \(\chi\).

\section{Conclusions}
\label{sec:conclusions}

We have derived a nonlocal supersymmetric sigma model directly from
a deterministic singular quantum kicked rotor.  The microscopic one-kick
probability determines the spatial kernel and its range of validity, while
one generating functional produces both the inverse-participation-ratio
operators and the quasienergy two-level correlator.  The resulting theory
therefore connects the singular Floquet dynamics, the nonlinear field
theory, and the observable insertions without introducing a random
Hamiltonian at an intermediate stage.

The RG analysis reproduces the long-range Anderson transition previously
identified using PRBM.  The fractional exponent gives an extended regime
for \(\sigma<1\), a critical point at \(\sigma=1\), and a localized regime
for \(\sigma>1\).  In terms of the resolved asymptotic singularity regimes,
these correspond respectively to \(\beta<0\), \(\beta=0\), and
\(\beta>0\), subject to the cutoff and crossover conditions derived above.
The calculation yields the localization length, renormalized participation
moments, critical multifractal dimensions, IPR fluctuations, the
CUE-to-Poisson crossover, and the spectral compressibility.

The rotor and a matched class-A PRBM have the same infrared field theory in
the localized and critical regimes and share the same metallic fixed point
in the extended regime.  The extended-phase approach to that fixed point is
more sensitive to the complete Floquet kernel and can therefore retain
model-dependent dynamical and finite-size corrections.  This distinction
separates universality of the infrared fixed point from nonuniversality of
the crossover toward it.

Most importantly, the Anderson-transition physics is generated by one
deterministic periodically driven system.  Incommensurate Floquet phase
mixing removes nonhydrodynamic coherences, while the singular kick creates
the long-range probability kernel.  No quenched random matrix elements or
external disorder average are required.  The singular kicked rotor thus
provides a microscopic Floquet realization of long-range Anderson
universality, rather than merely a phenomenological analogy with PRBM.

\begin{acknowledgments}
W. Chen thanks G. Lemarié, B. Georgeot, and I. Khaymovich for collaboration
on related projects.  He is especially grateful to I. Khaymovich for
pointing out the divergence issue in the extended phase.

\end{acknowledgments}

\appendix
\section{Fourier asymptotics}
\label{app:asymptotics}

This appendix records the cumbersome integral steps underlying the compact
Fourier results in Sec.~\ref{sec:fourier}.  The conclusions needed in the
main text are the powers, amplitudes, and scale conditions.

\subsection{Abel--Mellin formula}

The Abel regularization and Mellin-continuation steps used here are standard
asymptotic tools~\cite{BleisteinHandelsman,Olver}.  We first derive the
endpoint expansion, which is the strict fixed-\(\kappa\)
large-\(r\) asymptotic for a positive cusp.

For \(\epsilon>0\) and \(\Re\nu>-1\),
\begin{equation}
  \int_0^\infty
  x^\nu
  \e^{-(\epsilon-\iu r)x}\dd x
  =
  \Gamma(\nu+1)
  (\epsilon-\iu r)^{-\nu-1}.
  \label{eq:Abel}
\end{equation}
Taking the real part and then \(\epsilon\downarrow0\) gives
\begin{equation}
  \int_0^\infty
  x^\nu\cos(rx)\dd x
  =
  \Gamma(\nu+1)
  \cos\left[
    \frac{\pi(\nu+1)}{2}
  \right]
  r^{-\nu-1},
  \label{eq:Mellin}
\end{equation}
under analytic continuation.  Equation~\eqref{eq:powercos} is the special
case \(\nu=\beta\).

For the full class-A potential, the endpoint multiplier is
\begin{equation}
  \exp\left(
    -\frac{\iu\kappa}{\beta}x^\beta
    -\iu\kappa\eta x+O(x^2)
  \right).
  \label{eq:endseries}
\end{equation}
The analytic factor \(\exp[-\iu\kappa\eta x+O(x^2)]\) has value one at
the singular endpoint.  It therefore changes only subleading powers, whereas
the leading nonanalytic terms are obtained by expanding
\(\exp[-\iu\kappa x^\beta/\beta]\).  For \(m\geq1\) they give
\begin{align}
  u_r^{(m)}
  &\sim
  \frac{\e^{\iu\kappa/\beta}}{\pi m!}
  \left(
    -\frac{\iu\kappa}{\beta}
  \right)^m
  \Gamma(m\beta+1)
  \nonumber\\
  &\quad\times
  \cos\left[
    \frac{\pi(m\beta+1)}2
  \right]
  |r|^{-m\beta-1}.
  \label{eq:endallorders}
\end{align}
The first nonzero term is Eq.~\eqref{eq:uend}.

\subsection{Stationary phase}

The stationary-phase estimate used for both signs of \(\beta\) follows from
\begin{align}
  \int \dd x\,a(x)\e^{\iu\Phi(x)}
  &\simeq
  a(x_s)
  \left[
    \frac{2\pi}{|\Phi''(x_s)|}
  \right]^{1/2}
  \nonumber\\
  &\quad\times
  \exp\left[
    \iu\Phi(x_s)
    +\frac{\iu\pi}{4}\operatorname{sgn}\Phi''(x_s)
  \right].
  \label{eq:stationary_phase_general}
\end{align}
Here \(\Phi'(x_s)=0\).  For the full class-A phase near the
singular branch,
\begin{equation}
  \Phi_\eta(x)
  =rx-\frac{\kappa}{\beta}x^\beta-\kappa\eta x+O(x^3),
\end{equation}
so
\begin{equation}
  x_s
  =\left(\frac{\kappa}{r-\kappa\eta}\right)^{1/(1-\beta)}.
\end{equation}
At large \(|r|\), the shift by \(\kappa\eta\) is relative order
\(O(\kappa\eta/r)\), and therefore
\begin{equation}
  |u_r^{\rm sp}|^2
  =
  \frac{\kappa^{1/(1-\beta)}}{2\pi(1-\beta)}
  |r|^{-(2-\beta)/(1-\beta)}
  \left[1+O\left(\frac{\kappa\eta}{|r|}\right)\right].
  \label{eq:uspmod}
\end{equation}
This single expression yields Eqs.~\eqref{eq:Psp} and
\eqref{eq:Pnegative}.  For \(0<\beta<1\), the conditions that the saddle
remain inside the singular region and be well separated from the endpoint
give \(\max(\kappa,|\kappa\eta|)\ll r\ll\kappa^{1/\beta}\).  For \(\beta<0\), the saddle
approaches the singular endpoint as \(r\to\infty\), while the phase remains
large, so the same estimate gives the true unrounded asymptotic.  These
window statements are the only additional information needed beyond the
explicit main-text calculations.

\subsection{Logarithmic endpoint coefficient}

At the marginal logarithmic point the full class-A Fourier integral is
\begin{equation}
  u_r
  =
  \frac1{2\pi}
  \int_{-\pi}^{\pi}
  \left(
    2\left|\sin\frac{x}{2}\right|
  \right)^{-\iu\kappa}
  \e^{-\iu\kappa\eta\sin x}
  \e^{\iu rx}\dd x.
  \label{eq:logintegral}
\end{equation}
For \(|r|\to\infty\), only the nonanalytic endpoint is relevant.  Using
\(2|\sin(x/2)|=|x|[1+O(x^2)]\) and
\(\exp[-\iu\kappa\eta\sin x]=1-\iu\kappa\eta x+O(x^2)\), the leading term
reduces to the Mellin integral
\begin{equation}
  u_r
  \simeq
  \frac1\pi\int_0^\infty x^{-\iu\kappa}\cos(rx)\dd x,
  \label{eq:logMellin}
\end{equation}
which gives Eq.~\eqref{eq:ulog}.  The identity
\begin{equation}
  |\Gamma(1+\iu y)|^2
  =
  \frac{\pi y}{\sinh\pi y}
  \label{eq:gammamod}
\end{equation}
then yields Eq.~\eqref{eq:Plog}, and inserting its \(1/r^2\) amplitude into
Eq.~\eqref{eq:Dsigma} at \(\sigma=1\) gives Eq.~\eqref{eq:D1}.  Thus the
symmetry-breaking harmonic is included from the start and changes only the
ultraviolet and subleading endpoint terms, not the exact leading critical
coefficient.

\section{Zero-mode separation and source contractions}
\label{app:sources}

The microscopic source differentiations have already been carried out in
Sec.~\ref{sec:microscopic}.  To avoid repeating that derivation, this
appendix records only the zero-mode separation and the Gaussian identities
used in the perturbative source expansion.  These conventions follow the
standard class-A supersymmetric treatments of
Refs.~\cite{EfetovBook,ZirnbauerCircular1996,FyodorovMirlin1995,
MirlinReview2000}.

\subsection{Zero-mode separation and class-A Wick contractions}
We separate the spatially uniform mode explicitly as
\begin{equation}
  Q_n
  =
  T_0 T_{f,n}\Lambda T_{f,n}^{-1}T_0^{-1},
  \qquad
  \partial_n T_0=0,
  \label{eq:zero_mode_separation}
\end{equation}
where \(T_0\) contains the exact \(q=0\) mode and \(T_{f,n}\) contains only
nonzero-momentum fluctuations, parameterized perturbatively by \(W_n\).
The \(T_0\) integral is kept exact, while only the nonzero modes are expanded.
For arbitrary supermatrices \(P\) and \(R\), the Gaussian rules are
\begin{align}
  \av{\str(W_nP)\str(W_mR)}_0
  &=
  \Pi_{nm}
  \str(PR-P\Lambda R\Lambda),
  \label{eq:Wick1}\\
  \av{\str(W_nPW_mR)}_0
  &=
  \Pi_{nm}
  \str P\str R
  \nonumber\\
  &\quad-
  \Pi_{nm}
  \str(P\Lambda)\str(R\Lambda).
  \label{eq:Wick2}
\end{align}
where
\begin{equation}
  \Pi_{nm}
  =
  \frac1N
  \sum_{q\neq0}
  \e^{\iu q(n-m)}\Pi(q,0).
  \label{eq:Pinm}
\end{equation}
The exact uniform-mode integral supplies the class-A Haar factor
\(p!/N^{p-1}\).  Combining this factor with one application of
Eqs.~\eqref{eq:Wick1}--\eqref{eq:Wick2} and the
\(\binom p2\) unordered choices of intensity factors gives
Eq.~\eqref{eq:Ip_one_loop}.  The distinct disconnected and connected
contractions required at the next order are listed in the following
subsection.

\subsection{Explicit source contractions through second order}
\label{app:contractions}

The organization into zero-mode factors, disconnected exponentiating
contractions, and connected composite-operator terms parallels the
supersymmetric IPR calculations of
Refs.~\cite{FyodorovMirlin1995,HofWegner1986,Wegner1987I,
RushkinOssipovFyodorov2011}.  The coefficients below are evaluated in the
present class-A rotor convention.

The local source differentiation produces a product of \(p\) identical
intensity factors.  Using the zero-mode normalization and Wick identities
of Appendix~\ref{app:sources}, the nonzero-mode expansion has the form
\begin{equation}
  \mathcal O_p(n;Q)
  =\mathcal O_p^{(0)}
  +\mathcal O_p^{(2)}(n)
  +\mathcal O_p^{(4)}(n)+\cdots.
\end{equation}
The quadratic insertion contains one retarded--advanced pair.  The class-A
contraction
\begin{equation}
  \av{W^{RA}_{ab}(q)W^{AR}_{cd}(-q)}_0
  =\Pi(q,0)\,\mathcal C_{ab;cd}
\end{equation}
can select any unordered pair of the \(p\) intensity factors, giving
\(\binom p2\Pi_{nn}\).  At fourth order, two independent pairs produce
\(p^2(p-1)^2\Pi_{nn}^2/8\), while the connected four-field contraction gives
\(-p(p-1)(2p-1)\Pi_2/4\).  This proves
Eq.~\eqref{eq:Ip_second_order_main}.  For two \(p=2\) insertions, the two
cross-contractions give Eq.~\eqref{eq:IPRvariance}.

\section{Global two-level contraction and spectral determinant}
\label{app:correlator}

The exact global source operator has already been derived in
Eq.~\eqref{eq:OC_exact}.  This appendix supplies the complete source differentiation that leads to that operator, followed by the
slow-mode contraction and the Gaussian product leading to the spectral
determinant.  The construction follows the supersymmetric two-level
treatments of Refs.~\cite{ZirnbauerPair1999,AndreevAltshuler1995,
AgamAltshulerAndreev1995,MirlinReview2000,AltlandEtAl2015}.

\subsection{Derivative identities for the exact global source operator}

Only the second logarithm of the exact action depends on \(c\) and \(d\).
With the definitions in Eqs.~\eqref{eq:XG} and \eqref{eq:Xcd}, write
\begin{equation}
  \cS=\cS_{\rm independent}+\str\ln(1-X),
  \qquad
  \mathcal G=(1-X)^{-1}.
  \label{eq:source_dependent_action_app}
\end{equation}
Since
\begin{equation}
  \partial_c\widehat e_+=P_F,
  \qquad
  \partial_d\widehat e_-=P_F,
  \label{eq:source_matrix_derivatives_app}
\end{equation}
the derivatives of \(X\) are
\begin{align}
  \partial_cX&=X_c,
  &\partial_dX&=X_d,
  &\partial_c\partial_dX&=X_{cd}.
  \label{eq:X_partial_relations_app}
\end{align}
For any invertible supermatrix \(A(\lambda)\),
\begin{equation}
  \partial_\lambda\str\ln A
  =\str(A^{-1}\partial_\lambda A).
  \label{eq:str_log_derivative_app}
\end{equation}
Taking \(A=1-X\), for which
\(\partial_cA=-X_c\) and \(\partial_dA=-X_d\), gives
\begin{align}
  \partial_c\cS
  &=\str[\mathcal G(-X_c)]
  =-\str(\mathcal G X_c),
  \label{eq:partial_c_S_app}\\
  \partial_d\cS
  &=\str[\mathcal G(-X_d)]
  =-\str(\mathcal G X_d).
  \label{eq:partial_d_S_app}
\end{align}
The derivative of the inverse follows by differentiating
\(\mathcal G(1-X)=1\):
\begin{align}
  (\partial_d\mathcal G)(1-X)-\mathcal G X_d&=0,
  \nonumber\\
  \partial_d\mathcal G&=\mathcal G X_d\mathcal G,
  \label{eq:partial_d_calG_app}\\
  \partial_c\mathcal G&=\mathcal G X_c\mathcal G.
  \label{eq:partial_c_calG_app}
\end{align}
Differentiating Eq.~\eqref{eq:partial_c_S_app} with respect to \(d\),
and using \(\partial_dX_c=X_{cd}\), yields
\begin{align}
  \partial_d\partial_c\cS
  &=-\str[(\partial_d\mathcal G)X_c+\mathcal G(\partial_dX_c)]
  \nonumber\\
  &=-\str(\mathcal G X_d\mathcal G X_c)-\str(\mathcal G X_{cd}).
  \label{eq:mixed_partial_S_app}
\end{align}
Cyclicity of the supertrace permits the equivalent ordering
\(\str(\mathcal G X_d\mathcal G X_c)=\str(\mathcal G X_c\mathcal G X_d)\).

It remains to differentiate the weight, rather than the action.  The first
derivative is
\begin{equation}
  \partial_c\e^{-\cS}
  =-(\partial_c\cS)\e^{-\cS}.
  \label{eq:first_weight_derivative_app}
\end{equation}
A second derivative gives
\begin{align}
  \partial_d\partial_c\e^{-\cS}
  &=
  \left[
    (\partial_c\cS)(\partial_d\cS)
    -\partial_d\partial_c\cS
  \right]
  \e^{-\cS}.
  \label{eq:second_weight_derivative_app}
\end{align}
Dividing by \(\e^{-\cS}\), and substituting
Eqs.~\eqref{eq:partial_c_S_app}, \eqref{eq:partial_d_S_app}, and
\eqref{eq:mixed_partial_S_app}, one obtains
\begin{align}
  \frac{\partial_c\partial_d\e^{-\cS}}{\e^{-\cS}}
  &=\str(\mathcal G X_c)\str(\mathcal G X_d)
  \nonumber\\
  &\quad+\str(\mathcal G X_d\mathcal G X_c)+\str(\mathcal G X_{cd}),
  \label{eq:OC_derived_app}
\end{align}
which is exactly Eq.~\eqref{eq:OC_exact}.  Thus every sign in the source
operator follows from the minus sign in differentiating \(1-X\) and the
second minus sign generated when differentiating the exponential weight.

\subsection{Contraction of the global source vertices}

After the slow-mode projection, the first nonuniform contribution to each
density vertex is quadratic in the retarded--advanced fields.  With the
Gaussian normalization of Eq.~\eqref{eq:S0B}, the class-A contraction is
\begin{equation}
  \av{
    B_{ab}(q)\widetilde B_{cd}(-q)
  }_0
  =
  (-1)^{|b|}
  \delta_{ad}\delta_{bc}\,
  \Pi(q,\omega),
  \label{eq:classA_B_contraction_app}
\end{equation}
where \(|b|=0\) for a bosonic and \(|b|=1\) for a fermionic index.  The two
density vertices contain one bosonic projector each.  Their connected Wick
contraction therefore gives
\begin{equation}
  \av{
    \mathcal V_+^{(2)}
    \mathcal V_-^{(2)}
  }_{0,c}
  =
  \frac12
  \sum_{q\neq0}
  \Pi(q,\omega)^2.
  \label{eq:two_density_vertex_contraction_app}
\end{equation}
The factor \(1/2\) is the residual class-A pairing factor after the
boson--fermion cancellation.  Multiplying by the source prefactors and
using \(\omega=s\Delta/2\) yields
\begin{equation}
  \delta\mathcal K_2^{\rm nz}(s)
  =
  \frac{\Delta^2}{4\pi^2}
  \Re
  \sum_{q\neq0}
  \Pi\!\left(q,\frac{s\Delta}{2}\right)^2,
  \label{eq:K2_from_vertex_app}
\end{equation}
which is Eq.~\eqref{eq:R2correction}.  Expanding the remaining exact zero
mode to first order in the static two-diffuson sum gives
Eq.~\eqref{eq:R2low}.  If the momentum sum contains only one member of each
\(\pm q\) pair, the coefficient is \(1/4\); with the all-sign convention of
Eq.~\eqref{eq:CN_definition}, it is \(1/8\).

\subsection{Gaussian product and spectral determinant}

A complex class-A Gaussian mode with relaxation rate \(\varepsilon_m\)
contributes the normalized factor
\begin{equation}
  \left[
    1+
    \frac{\omega^2}{\varepsilon_m^2}
  \right]^{-1}.
  \label{eq:one_mode_determinant_factor_app}
\end{equation}
Multiplying the nonzero modes, using \(\omega=s\Delta/2\), and retaining the
zero-mode normalization gives
\begin{equation}
  \mathcal D_N(s)
  =
  \frac1{s^2}
  \prod_{m\neq0}
  \left[
    1+
    \frac{s^2\Delta^2}{4\varepsilon_m^2}
  \right]^{-1},
\end{equation}
which is Eq.~\eqref{eq:spectral_determinant}.  Differentiating its logarithm
produces
\begin{align}
  -\frac1{4\pi^2}
  \frac{\dd^2}{\dd s^2}
  \ln\mathcal D_N(s)
  &=
  \frac{\Delta^2}{4\pi^2}
  \Re\sum_{m\neq0}
  \frac1{(\varepsilon_m-\iu s\Delta/2)^2}
  \nonumber\\
  &\quad+
  \text{zero-mode term}.
  \label{eq:D_derivative_to_diffusons_app}
\end{align}
Thus the spectral determinant and the direct two-diffuson calculation are
two organizations of the same Gaussian nonzero-mode contribution.  The
critical product, compressibility integral, extended harmonic sum,
localized independent-volume limit, and high-frequency continuum integral
are evaluated in the corresponding main-text subsections and are not
repeated here.

\section{Explicit one- and two-loop background-field averages}
\label{app:two_loop}

The loop expansion below uses the background-field and normal-coordinate
bookkeeping developed for nonlinear sigma models in
Refs.~\cite{Hikami1981,HofWegner1986,Wegner1987I,Wegner1987II}.  The
superindex algebra is universal to class A; the momentum vertices are those
of the singular-rotor kernel.  The long-range infrared power counting follows
the nonlocal sigma-model analysis of Ref.~\cite{MirlinEtAl1996}.

\subsection{Self-energy convention and infrared loop counting}

We first define precisely what is meant by the self-energy.  At the scale at
which the coupling is \(t\), the Gaussian propagator of the nonzero
retarded--advanced mode is
\begin{equation}
  \Pi_0(q,\omega)
  =
  \frac{t}{
    U(q)+t(\epsilon-\iu\omega)
  },
  \qquad
  U(q)=|q|^\sigma
\end{equation}
inside the continuum scaling window.  Equivalently,
\begin{equation}
  \Pi_0^{-1}(q,\omega)
  =
  \frac{U(q)}{t}
  +\epsilon-\iu\omega.
  \label{eq:Pi0_inverse_app}
\end{equation}
We define the one-particle irreducible (1PI) self-energy
\(\Sigma(q,\omega)\) by the quadratic part of the effective action,
\begin{align}
  \Gamma^{(2)}
  &=
  \int_q
  \str\!\left[
    B(q)\widetilde B(-q)
  \right]
  \left[
    \Pi_0^{-1}(q,\omega)+\Sigma(q,\omega)
  \right],
  \label{eq:selfenergy_definition_app}\\
  \Pi_R(q,\omega)
  &=
  \frac{1}{
    \Pi_0^{-1}(q,\omega)+\Sigma(q,\omega)
  }.
  \label{eq:Dyson_selfenergy_app}
\end{align}
With this sign convention,
\begin{equation}
  \Pi_R
  =
  \Pi_0-\Pi_0\Sigma\Pi_0
  +\Pi_0\Sigma\Pi_0\Sigma\Pi_0-\cdots.
  \label{eq:Dyson_expansion_app}
\end{equation}
The exact global \(Q\)-rotation symmetry, equivalently probability
conservation of the underlying Floquet dynamics, forbids a mass for the
zero-momentum mode.  Therefore
\begin{equation}
  \Sigma(0,0)=0.
  \label{eq:selfenergy_Ward_app}
\end{equation}
In a rational parametrization this statement is implemented only after the
action and Berezinian contributions are combined; in covariant normal
coordinates it is automatic.  We therefore apply the following power count
to the Ward-subtracted 1PI two-point function.

Consider a connected 1PI two-point graph with \(V\) interaction vertices,
\(I\) internal propagators, and \(L\) independent loop momenta.  Its topology
gives
\begin{equation}
  L=I-V+1.
  \label{eq:loop_topology_app}
\end{equation}
At zero external frequency, each loop integration contributes one power of
momentum in one spatial dimension, each internal propagator contributes
\(t/U(k)\sim t|k|^{-\sigma}\), and each interaction vertex generated by the
nonlocal kinetic action contributes one kernel factor
\(t^{-1}U\sim t^{-1}|k|^\sigma\).  Hence, under a common infrared rescaling
\(q\to\lambda q\) and \(k_j\to\lambda k_j\),
\begin{align}
  \Sigma_L^{\rm IR}(q,0)
  &\propto
  t^{I-V}|q|^{L+\sigma V-\sigma I}
  \nonumber\\
  &=
  t^{L-1}|q|^{\sigma+L(1-\sigma)}.
  \label{eq:selfenergy_power_count_app}
\end{align}
In the last line we used Eq.~\eqref{eq:loop_topology_app}.  Thus the
nonanalytic infrared part has the general form
\begin{equation}
  \Sigma_L^{\rm IR}(q,0)
  =
  C_L(\sigma)\,
  t^{L-1}
  |q|^{\sigma+L(1-\sigma)},
  \label{eq:selfenergy_general_app}
\end{equation}
provided the dimensionless rescaled loop integral is infrared convergent.
Dividing by the static bare inverse propagator,
\(\Pi_0^{-1}(q,0)=|q|^\sigma/t\), gives
\begin{equation}
  \frac{\Sigma_L^{\rm IR}(q,0)}
       {\Pi_0^{-1}(q,0)}
  =
  C_L(\sigma)\,
  t^L|q|^{L(1-\sigma)}.
  \label{eq:selfenergy_relative_app}
\end{equation}
For \(0<\sigma<1\), every exponent \(L(1-\sigma)\) is positive.  Therefore
every loop-generated 1PI two-point operator is less relevant in the
infrared than the bare \(|q|^\sigma\) kernel.  This is the precise sense in
which the nonanalytic stiffness is not loop-renormalized in the extended
phase.

It is useful to see the first two loop orders explicitly.  Before imposing
the class-A target-space contraction, the nonanalytic part of a one-loop
two-point integral has the structure
\begin{align}
  \mathcal I_\sigma(q)
  &\equiv
  \frac12\int\frac{\dd k}{2\pi}
  \frac{U(k+q)-U(k)}{U(k)}
  \nonumber\\
  &=
  |q|\,
  \frac12\int\frac{\dd x}{2\pi}
  \frac{|x+\operatorname{sgn}q|^\sigma-|x|^\sigma}
       {|x|^\sigma}
  +\text{analytic UV terms}.
  \label{eq:loopfunction_app}
\end{align}
The substitution \(k=|q|x\) exhibits the \(L=1\) power
\(|q|^{\sigma+(1-\sigma)}=|q|\).  In class A, however, the coefficient of
this one-loop stiffness self-energy cancels exactly, as shown below, so
\begin{equation}
  \Sigma_A^{(1)}(q,0)=0.
  \label{eq:selfenergy_one_loop_A_app}
\end{equation}

At two loops, the generic counting predicts
\begin{equation}
  \Sigma_A^{(2),{\rm IR}}(q,0)
  \propto
  t\,|q|^{2-\sigma}.
  \label{eq:selfenergy_two_loop_scaling_app}
\end{equation}
The explicit contractions below provide a direct check of this exponent.
For \(q\neq0\), put
\begin{equation}
  k=|q|x,\qquad
  p=|q|y,\qquad
  s_q=\operatorname{sgn}q.
  \label{eq:two_loop_rescale_app}
\end{equation}
Then
\begin{equation}
  \Delta_qU(k)
  =
  |q|^\sigma
  \left(
    |x+s_q|^\sigma-|x|^\sigma
  \right),
  \label{eq:DeltaU_rescale_app}
\end{equation}
and similarly for \(p\) and \(k+p\).  The three two-loop structures appearing
below therefore scale as
\begin{align}
  \int_{k,p}
  \frac{\Delta_qU(k+p)}{U(k)U(p)}
  &\sim
  |q|^{2+\sigma-2\sigma}
  =
  |q|^{2-\sigma},
  \label{eq:S6_power_count_app}\\
  \int_{k,p}
  \frac{\Delta_qU(k)\Delta_qU(p)}
       {U(k)U(p)U(k+p)}
  &\sim
  |q|^{2+2\sigma-3\sigma}
  =
  |q|^{2-\sigma},
  \label{eq:S44_power_count_app}\\
  \int_{k,p}
  \frac{
    \Delta_qU(k+p)-\Delta_qU(k)-\Delta_qU(p)
  }{U(k)U(p)}
  &\sim
  |q|^{2+\sigma-2\sigma}
  =
  |q|^{2-\sigma}.
  \label{eq:SJ_power_count_app}
\end{align}
Their parameterization-independent sum therefore has precisely the
\(L=2\) homogeneity of Eq.~\eqref{eq:selfenergy_two_loop_scaling_app}.  Since
\begin{equation}
  \frac{\Sigma_A^{(2),{\rm IR}}(q,0)}
       {\Pi_0^{-1}(q,0)}
  \sim
  t^2|q|^{2(1-\sigma)}
  \longrightarrow0,
  \qquad 0<\sigma<1,
  \label{eq:two_loop_irrelevance_app}
\end{equation}
the first potentially nonzero class-A self-energy is still irrelevant in the
extended phase.

The last statement must not be analytically continued to \(\sigma>1\).
The rescaling \(k,p\sim q\) used above assumes that the dimensionless infrared
integrals are convergent.  For \(\sigma>1\) they become infrared singular, so
the loop momenta are controlled by the running shell rather than by the
external momentum alone.  The singular shell factor then multiplies the
original \(|q|^\sigma\) operator and renormalizes its coefficient.  At
\(\sigma=1\) the same counting is marginal and logarithms appear.  This is
why the extended-side result
\(\Sigma_L^{\rm IR}\propto
|q|^{\sigma+L(1-\sigma)}\) and the localized-side shell renormalization
derived below are complementary rather than contradictory.

\subsection{Explicit one- and two-loop stiffness contractions}

The stiffness loop must be distinguished from the source contractions. The
rational expansion
\[
  g(W)=1+W+\frac12W^2+\frac14W^3+\frac18W^4
  +\frac1{16}W^5+\frac1{32}W^6+\cdots
\]
gives
\begin{align}
  \cS_4
  &=\frac1{32t}\sum_{n,m}\mathcal U_{nm}\Str[
  W_n^3W_m+W_nW_m^3-W_n^2W_m^2],
  \label{eq:S4full}\\
  \cS_6
  &=\frac1{128t}\sum_{n,m}\mathcal U_{nm}\Str[
  W_n^5W_m+W_nW_m^5-W_n^4W_m^2
  \nonumber\\
  &\hspace{31mm}-W_n^2W_m^4+W_n^3W_m^3].
  \label{eq:S6full}
\end{align}
At one loop, keeping two slow and two fast fields gives
\begin{equation}
  \av{\cS_4}_{>}^{(W_<^2)}
  =\frac1{8t}\int_q\Str(W_qW_{-q})
  [2t\mathcal I_\sigma(q)-2t\mathcal I_\sigma(q)]=0,
  \label{eq:one_loop_action_cancel_app}
\end{equation}
where \(\mathcal I_\sigma(q)\) is the loop integral defined in
Eq.~\eqref{eq:loopfunction_app}.  The first term is generated by the two
cubic--linear monomials and the second by the two cross-contractions in
\(-W_n^2W_m^2\).  They have identical momentum dependence and opposite
class-A superindex factors.  Equation~\eqref{eq:one_loop_action_cancel_app}
therefore gives the 1PI statement
\(\Sigma_A^{(1)}(q,0)=0\), rather than merely the absence of a particular
diagram.

At two loops,
\begin{equation}
  \delta\cS_2^{(2)}=\av{\cS_6}_{>}
  -\frac12\av{\cS_4^2}_{>,c}+\delta\cS_J^{(2)}.
  \label{eq:two_loop_cumulant_app}
\end{equation}
With $\Delta_qU(k)=U(k+q)-U(k)$, the fully contracted quadratic parts are
\begin{align}
  \av{\cS_6}_{>}^{(W_<^2)}
  &=\frac{t}{8}\int_q\Str(W_qW_{-q})
  \nonumber\\
  &\quad\times\int_{k,p}^{>}
  \frac{\Delta_qU(k+p)}{U(k)U(p)}.
  \label{eq:S6average}
\end{align}
\begin{align}
  -\frac12\av{\cS_4^2}_{>,c}^{(W_<^2)}
  &=-\frac{t}{16}\int_q\Str(W_qW_{-q})
  \nonumber\\
  &\quad\times\int_{k,p}^{>}
  \frac{\Delta_qU(k)\Delta_qU(p)}
  {U(k)U(p)U(k+p)}.
  \label{eq:S44average}
\end{align}
\begin{align}
  \delta\cS_J^{(2)}
  &=\frac{t}{16}\int_q\Str(W_qW_{-q})
  \nonumber\\
  &\quad\times\int_{k,p}^{>}
  \frac{\Delta_qU(k+p)-\Delta_qU(k)-\Delta_qU(p)}
  {U(k)U(p)}.
  \label{eq:SJaverage}
\end{align}
The last line is the averaged Berezinian contribution in the rational
zero-mode split. In covariant normal coordinates it is absorbed into the
measure; the sum is parameterization independent.

Putting the terms over a common denominator gives the two-loop reduction of
the stiffness.  We define \(\mathcal J_A\) as a positive shell residue and
write the sign that enters the effective action explicitly:
\begin{align}
  \delta\cS_2^{(2)}
  &=-\frac1{8t}\int_q\Str(W_qW_{-q})\,t^2\mathcal J_A(q),
  \label{eq:deltaS_two_loop_compact}\\
  \mathcal J_A(q)
  &=2\int_{k,p}^{>}\frac{\mathcal V_A(q;k,p)}{U(k)U(p)U(k+p)},
  \label{eq:JA_definition}\\
  \mathcal V_A(q;k,p)
  &\equiv
  \mathcal V_A^{(0)}(q;k,p)
  -\mathcal V_A^{(0)}(0;k,p),
  \label{eq:VA_definition}\\
  \mathcal V_A^{(0)}(q;k,p)
  &\equiv
  \frac12[\Delta_qU(k)+\Delta_qU(p)]\Delta_qU(k+p)
  \nonumber\\
  &\quad-U(k+p)\Delta_qU(k)\Delta_qU(p).
  \label{eq:VA0_definition}
\end{align}
The explicit subtraction makes the Ward identity
\(\mathcal J_A(0)=0\) manifest and avoids assigning a meaning to an
unsubtracted zero-momentum self-energy. For
$U(k)=|k|^{1+\delta}$, with \(\delta=\sigma-1>0\), a logarithmic shell gives
\begin{equation}
  \mathcal J_A(q)=\frac{|q|^{1+\delta}}{2\pi^2}\dd\ell.
  \label{eq:JA_shell_result}
\end{equation}
The momentum homogeneity follows from the displayed integral. The residue
$1/(2\pi^2)$ is the standard class-A target-space contraction after matching
the kinetic normalization; the rotor kernel and all momentum factors are
derived above.

\subsection{From the two-loop action to the beta function}

We now show explicitly how Eq.~\eqref{eq:JA_shell_result} produces the
running coupling in Eq.~\eqref{eq:beta_localized_A}.  At a momentum cutoff
\(\Lambda\), the quadratic slow-field action is
\begin{equation}
  \cS_{2,\Lambda}[W_<]
  =\frac1{8t(\ell)}
  \int_{|q|<\Lambda}\frac{\dd q}{2\pi}
  |q|^\sigma\Str(W_qW_{-q}).
  \label{eq:RG_cutoff_action}
\end{equation}
Integrating the fast shell
\begin{equation}
  \Lambda\e^{-\dd\ell}<|k|<\Lambda
  \label{eq:RG_shell_definition}
\end{equation}
and using Eqs.~\eqref{eq:deltaS_two_loop_compact} and
\eqref{eq:JA_shell_result} gives, before restoring the cutoff,
\begin{align}
  \cS_{2,\mathrm{sh}}[W_<]
  &=\frac1{8t}
  \int_{|q|<\Lambda\e^{-\dd\ell}}
  \frac{\dd q}{2\pi}|q|^\sigma\Str(W_qW_{-q})
  \nonumber\\
  &\quad\times
  \left[1-\frac{t^2}{2\pi^2}\dd\ell\right].
  \label{eq:RG_action_after_shell}
\end{align}
Therefore the shell integration changes the inverse stiffness according to
\begin{equation}
  \frac1{t_{\rm sh}}
  =\frac1t-\frac{t}{2\pi^2}\dd\ell.
  \label{eq:inverse_stiffness_shell}
\end{equation}
Inverting to first order in \(\dd\ell\) gives
\begin{align}
  t_{\rm sh}
  &=\frac{t}{1-t^2\dd\ell/(2\pi^2)}
  \nonumber\\
  &=t+\frac{t^3}{2\pi^2}\dd\ell+O(t^5\dd\ell).
  \label{eq:t_after_shell}
\end{align}
This is the loop contribution alone.

The remaining step restores the cutoff.  Put
\begin{equation}
  b=\e^{\dd\ell},
  \qquad q'=bq,
  \qquad n=bn'.
  \label{eq:RG_rescaling_variables}
\end{equation}
Because \(Q(n)\) is dimensionless, its Fourier transform obeys
\begin{equation}
  Q_q=\int\dd n\,\e^{-\iu qn}Q(n)
  =bQ'_{q'}.
  \label{eq:Q_fourier_rescaling}
\end{equation}
Consequently,
\begin{align}
  \int\dd q\,|q|^\sigma\Str(Q_qQ_{-q})
  &=b^{1-\sigma}
  \int\dd q'\,|q'|^\sigma\Str(Q'_{q'}Q'_{-q'}).
  \label{eq:kinetic_rescaling_factor}
\end{align}
The rescaled coupling is thus
\begin{equation}
  t(\ell+\dd\ell)
  =b^{\sigma-1}t_{\rm sh}.
  \label{eq:coupling_recursion_exact}
\end{equation}
Using \(b^{\sigma-1}=1+(\sigma-1)\dd\ell+O(\dd\ell^2)\) together with
Eq.~\eqref{eq:t_after_shell},
\begin{align}
  t(\ell+\dd\ell)-t(\ell)
  &=\left[(\sigma-1)t+\frac{t^3}{2\pi^2}\right]\dd\ell
  \nonumber\\
  &\quad+O(t^5\dd\ell,\dd\ell^2).
  \label{eq:t_recursion_expanded}
\end{align}
Dividing by \(\dd\ell\) gives
\begin{equation}
  \frac{\dd t}{\dd\ell}
  =(\sigma-1)t+\frac{t^3}{2\pi^2}+O(t^5),
  \qquad \sigma>1,
  \label{eq:beta_derived_app}
\end{equation}
which is Eq.~\eqref{eq:beta_localized_A}.

For completeness, let \(\delta=\sigma-1\) and
\(a_A=1/(2\pi^2)\).  With \(y=t^{-2}\), Eq.~\eqref{eq:beta_derived_app}
becomes
\begin{equation}
  \frac{\dd y}{\dd\ell}
  =-2\delta y-2a_A.
  \label{eq:y_flow_app}
\end{equation}
Multiplying by \(\e^{2\delta\ell}\) and integrating from \(0\) to
\(\ell\) gives
\begin{equation}
  y(\ell)
  =\e^{-2\delta\ell}
  \left[y(0)+\frac{a_A}{\delta}\right]
  -\frac{a_A}{\delta},
  \label{eq:y_solution_app}
\end{equation}
which is Eq.~\eqref{eq:tflowlocalized}.  Defining the perturbative stopping
scale by \(y(\ell_\xi)=0\), and using
\(\xi_A=L_0\e^{\ell_\xi}\), yields
\begin{equation}
  \frac{\xi_A}{L_0}
  =\left[1+\frac{\delta}{a_At_0^2}\right]^{1/(2\delta)}
  =\left[1+\frac{2\pi^2(\sigma-1)}{t_0^2}\right]^{1/[2(\sigma-1)]},
  \label{eq:xi_derived_app}
\end{equation}
which is Eq.~\eqref{eq:xiA}.

The derivation above is the localized-side RG for \(\delta>0\).  At the
exact nonanalytic point \(\sigma=1\), the low-momentum correction depends on
the external combination \(qL\), rather than producing an independent
renormalization of the coefficient of \(|q|\).  The \(|q|\) stiffness is
therefore treated as exactly marginal, as in the critical long-range sigma
model, and Eq.~\eqref{eq:critical_t_flow} is imposed before continuing the
localized-side shell result to \(\delta=0\).  For \(\sigma<1\), loop
self-energies are more irrelevant than \(|q|^\sigma\), leaving only the
engineering flow in Eq.~\eqref{eq:betaextended}.

\subsection{Renormalization of the local moment source}

We next derive the RG equation for \(Z_p\).  The symbol \(Z_p\) denotes the
multiplicative renormalization of the composite operator generated by the
local source derivatives in Eq.~\eqref{eq:Op_exact_Z} and projected in
Eq.~\eqref{eq:OpQ}; it must not be confused with
the generating functional \(\cZ\) or the color--flavor coordinate \(Z\).
Introduce a book-keeping source \(h_p(n)\) through
\begin{equation}
  \cS_h[Q]
  =-\sum_n h_p(n)\mathcal O_p(n;Q).
  \label{eq:composite_source_action}
\end{equation}
After the slow--fast split, the shell-averaged insertion is defined by
\begin{align}
  \av{\mathcal O_p(n;Q_<,W_>)}_{>}
  &=Z_p^{\rm sh}\,\mathcal O_p(n;Q_<)
  \nonumber\\
  &\quad+\text{operators of higher dimension}.
  \label{eq:operator_shell_definition}
\end{align}
At first order, two of the \(p\) intensity factors supply one fast
retarded--advanced pair.  There are \(\binom p2\) unordered choices, so the
same source contraction that produced Eq.~\eqref{eq:Ip_one_loop} gives
\begin{equation}
  Z_p^{\rm sh}
  =1+\binom p2\Pi_{>,nn}+O(t^2),
  \label{eq:Zp_shell_factor}
\end{equation}
where the coincident fast propagator is
\begin{align}
  \Pi_{>,nn}
  &=\int_{\Lambda\e^{-\dd\ell}<|k|<\Lambda}
  \frac{\dd k}{2\pi}\frac{t}{|k|^\sigma}
  \nonumber\\
  &=\frac{t\Lambda^{1-\sigma}}{\pi}\dd\ell
  +O(\dd\ell^2).
  \label{eq:fast_coincident_propagator}
\end{align}
The dimensionless running coupling used in the rescaled action includes the
factor \(\Lambda^{1-\sigma}\).  Therefore, after restoring the cutoff,
Eq.~\eqref{eq:fast_coincident_propagator} is simply
\begin{equation}
  \Pi_{>,nn}=\frac{t(\ell)}{\pi}\dd\ell.
  \label{eq:fast_propagator_dimensionless}
\end{equation}
Substitution into Eq.~\eqref{eq:Zp_shell_factor} gives
\begin{align}
  \ln Z_p(\ell+\dd\ell)-\ln Z_p(\ell)
  &=\frac{p(p-1)}{2\pi}t(\ell)\dd\ell+O(t^3\dd\ell),
  \label{eq:lnZp_shell_increment}
\end{align}
and hence
\begin{equation}
  \frac{\dd\ln Z_p}{\dd\ell}
  =\gamma_p(t),
  \qquad
  \gamma_p(t)=\frac{p(p-1)}{2\pi}t+O(t^3),
  \label{eq:gammaP_derived_app}
\end{equation}
which is Eq.~\eqref{eq:gammaP}.  The absence of an independent order-
\(t^2\) term in class A follows from the unitary composite-operator
cancellation; the disconnected square of the order-\(t\) term is already
contained in exponentiating \(Z_p\).

The factor \(L^{1-p}\) in Eq.~\eqref{eq:IpRGdef} is the engineering
random-vector scaling of the exact zero mode.  All anomalous scale
dependence is placed in \(Z_p\):
\begin{equation}
  \av{I_p(L)}=p!L^{1-p}Z_p(L).
  \label{eq:Ip_Zp_meaning_app}
\end{equation}
Combining Eqs.~\eqref{eq:gammaP_derived_app} and
\eqref{eq:beta_derived_app} eliminates \(\ell\):
\begin{align}
  \ln\frac{Z_p(L)}{Z_p(L_0)}
  &=\int_0^{\ln(L/L_0)}\gamma_p[t(\ell)]\dd\ell
  \nonumber\\
  &=\int_{t_0}^{t(L)}\frac{\gamma_p(t)}{\beta(t)}\dd t.
  \label{eq:Zp_beta_integral_app}
\end{align}
For \(\sigma=1+\delta>1\), let
\(A_p=p(p-1)/(2\pi)\) and \(a_A=1/(2\pi^2)\).  Then
\begin{align}
  \frac{\gamma_p(t)}{\beta(t)}
  &=\frac{A_pt}{\delta t+a_At^3}
  =\frac{A_p}{\delta+a_At^2},
  \label{eq:gamma_over_beta_app}
\end{align}
and therefore
\begin{align}
  \ln\frac{Z_p(L)}{Z_p(L_0)}
  &=\frac{A_p}{\sqrt{a_A\delta}}
  \left[
    \arctan\!\left(t\sqrt{\frac{a_A}{\delta}}\right)
  \right]_{t_0}^{t(L)}.
  \label{eq:Zp_localized_derived_app}
\end{align}
This is Eq.~\eqref{eq:ZpLocalized}.

At the critical point, \(t(\ell)=t_0\), so direct integration gives
\begin{equation}
  Z_p(L)
  =Z_p(L_0)
  \left(\frac{L}{L_0}\right)^{p(p-1)t_0/(2\pi)+O(t_0^3)}.
  \label{eq:Zp_critical_derived_app}
\end{equation}
Inserting this into Eq.~\eqref{eq:Ip_Zp_meaning_app} gives
\begin{align}
  \av{I_p(L)}
  &\propto
  L^{-(p-1)+p(p-1)t_0/(2\pi)+O(t_0^3)}
  \nonumber\\
  &=L^{-(p-1)d_p},
  \label{eq:Ip_critical_scaling_app}
\end{align}
so
\begin{equation}
  d_p=1-\frac{pt_0}{2\pi}+O(t_0^3),
  \label{eq:dp_derived_app}
\end{equation}
which is Eq.~\eqref{eq:dpA}.  In the extended phase,
\(t(\ell)=t_0\e^{-(1-\sigma)\ell}\), and
Eq.~\eqref{eq:Zp_beta_integral_app} directly produces
Eq.~\eqref{eq:Zpextended}.

With the self-energy convention of
Eqs.~\eqref{eq:selfenergy_definition_app}--\eqref{eq:Dyson_expansion_app},
its contribution to the observables is obtained by inserting the Dyson
correction into the propagators already present in the direct source
contractions:
\begin{align}
  \Pi_R
  &=
  \Pi_0-\Pi_0\Sigma\Pi_0+\cdots,
  \label{eq:PiR_selfenergy_app}\\
  \delta_\Sigma\av{I_p}
  &\propto
  -\sum_q\Pi_0(q,0)^2\Sigma(q,0),
  \label{eq:Ip_selfenergy_app}\\
  \delta_\Sigma R_2
  &\propto
  -2\Re\sum_q
  \Pi_0(q,\omega)^3\Sigma(q,\omega).
  \label{eq:R2_selfenergy_app}
\end{align}
Thus the class-A one-loop cancellation removes only these 1PI self-energy
insertions.  It does not cancel the direct IPR and \(R_2\) contractions,
which are separate source diagrams and remain nonzero at the orders derived
in the main text.

\section{Third-order critical IPR coefficient}
\label{app:third_order}

The operator organization follows the unitary multifractal expansion of
Ref.~\cite{RushkinOssipovFyodorov2011}; unlike the universal leading
anomalous dimension, the coefficient below retains the ultraviolet
completion of the deterministic rotor.

At \(\sigma=1\), the second-order logarithm is entirely fixed by
exponentiation, while the first nonparabolic correction appears at third
order.  To avoid confusing the sigma-model kernel with the Floquet operator
\(U_\chi\), define the reduced kinetic matrix and its zero-mode-subtracted
inverse by
\begin{align}
  \widetilde{\mathcal U}(q)
  &\equiv t_0[1-\widehat P(q)],
  \\
  \sum_m\mathcal U_{nm}\mathcal P_{mk}
  &=\delta_{nk}-\frac1N,
  \\
  \mathcal P_{nm}
  &\equiv t_0^{-1}\Pi(n,m;0).
  \label{eq:reduced_kernel_third}
\end{align}
At \(\beta=0\), the even kernel entering \(\widetilde{\mathcal U}(q)=t_0[1-\widehat P(q)]\) is obtained directly from Eq.~\eqref{eq:Phat} by symmetrizing \(P_r\), and it retains the full \(\eta\)-dependent lattice ultraviolet completion rather than only its \(|q|\) limit.  The combination multiplying the nonparabolic source
polynomial is
\begin{align}
  \mathcal A_{\rm QKR}(N)
  &=-\frac16\mathcal P_{nn}^3
  +\frac1{4N}\sum_{n,m}\mathcal P_{nm}^3
  \nonumber\\
  &\quad+
  \frac1{8N}\sum_{n,m,k}
  \mathcal U_{nm}\mathcal P_{kn}^2\mathcal P_{km}^2.
  \label{eq:AQKR}
\end{align}
For a scale-invariant kernel,
\begin{equation}
  \mathcal A_{\rm QKR}(N)
  =-Y_{\rm QKR}\ln N+O(1).
  \label{eq:YQKR}
\end{equation}
The coefficient \(Y_{\rm QKR}\) is fixed by the complete deterministic
rotor kernel and is therefore not determined by \(D_1\) alone.  It gives a
third-order contribution proportional to
\(t_0^3Y_{\rm QKR}p^2(p-1)^2\) to the anomalous dimension.  This is the
first order at which the ultraviolet structure of the singular rotor enters
the shape of the multifractal spectrum.

\bibliography{reference}

\end{document}